\documentclass[12pt]{iopart}

\usepackage{iopams}  
\usepackage{graphicx} 
\usepackage{adjustbox}   

\begin{document}

\title{Geometric description of modular and weak values in discrete quantum systems using the Majorana representation}

\author{Mirko Cormann and Yves Caudano}

\address{Research Centre in Physics of Matter and Radiation (PMR) and Namur Center for Complex Systems (naXys), University of Namur, Rue de Bruxelles 61, B-5000 Namur, Belgium}
\ead{mirko.cormann@unamur.be, yves.caudano@unamur.be}
\begin{abstract}
We express modular and weak values of observables of three- and higher-level quantum systems in their polar form. The Majorana representation of $N$-level systems in terms of symmetric states of $N-1$ qubits provides us with a description on the Bloch sphere. With this geometric approach, we find that modular and weak values of observables of $N$-level quantum systems can be factored in $N-1$ contributions. Their modulus is determined by the product of $N-1$ ratios involving projection probabilities between qubits, while their argument is deduced from a sum of $N-1$ solid angles on the Bloch sphere. These theoretical results allow us to study the geometric origin of the quantum phase discontinuity around singularities of weak values in three-level systems. We also analyze the three-box paradox \cite{Aharonov (1991)} from the point of view of a bipartite quantum system. In the Majorana representation of this paradox, an observer comes to opposite conclusions about the entanglement state of the particles that were successfully pre- and postselected.
\end{abstract}

\maketitle

\section{Introduction}

Weak measurement experiments \cite{Aharonov (1988)} acquire limited information from a quantum system in order to preserve the coherence of the initial, preselected state $|\psi_{i}\rangle$ during its interaction with the meter system. After the probed system has undergone an additional postselection to the final state $|\psi_{f}\rangle$, the weak measurement outcome depends on an unbounded complex number called the weak value $A_{w}$:
\begin{equation}\label{eqWeakValueDef}
A_{w}=\frac{\langle\psi_{f}|\hat{A}|\psi_{i}\rangle}{\langle\psi_{f}|\psi_{i}\rangle}\:,
\end{equation}
where $\hat{A}$ is the observable studied during the weak measurement experiment. Weak values prove useful in precision metrology \cite{Hosten (2008),Dixon (2009),Starling (2009),Tang (2010),Rhee (2009)}, in the investigation of quantum paradoxes \cite{Resch (2004), George (2013), Lundeen (2009)}, in the reconstruction of initial qubit states \cite{Lundeen (2011),Salvail (2013),Lundeen (2012)}, or in the search for novel quantum effects \cite{Dreismann (2016)}. Modular values form a similar class of complex numbers describing pre- and postselected measurements of the observable $\hat{A}$. Modular values appear in quantum-gate type interactions \cite{Kedem (2010), Bin Ho (2016), Brun (2008)} and are the counterparts of weak values for unitary operators:
\begin{equation}\label{eqModularValueDef}
A_{m}=\frac{\langle\psi_{f}| e^{-j \theta \hat{A}}|\psi_{i}\rangle}{\langle\psi_{f}|\psi_{i}\rangle}\:,
\end{equation} 
where $\theta$ is a parameter representing an evolution strength and $j$ is the unit imaginary number. The hermitian operator $\hat{A}$ defines the unitary evolution $\hat{U}=e^{-j \theta \hat{A}}$. Modular values are not frequently reported explicitely in the literature because the standard weak measurement approximation links modular and weak values through a first-order polynomial development in the strength parameter $\theta$ \cite{Sponar (2015),Tollaksen (2010),Denkmayr (2014)}. Analytically, they are related through the exact expression $A_w=j  \left. {\frac{d A_m}{d \theta}  }\right\vert_{\theta = 0}$.

Weak and modular values of qubit observables feature a geometric representation in terms of three-dimensional vectors on the Bloch sphere \cite{Kofman (2012),Cormann (2015)}. The complex values are nicely expressed using their modulus and argument instead of their real and imaginary parts. The argument is connected to a solid angle: it depends on the area on the Bloch sphere surface enclosed during the state evolution from the initial preselected state to the final postselected state \cite{Cormann (2015)}. The argument has thus a topological origin, similar to the Pancharatnam geometric phase \cite{Pancharatnam (1956)}. This purely geometric approach is useful to understand rapid displacements of interference fringes in quantum eraser experiments \cite{Kobayashi (2011), Tamate (2009)}. It explains prior observations involving discontinuous phase jumps, such as the $\pi$-phase jump in cross-phase modulation \cite{Camacho (2009)}, as well as discontinuities around phase singularities \cite{Solli (2004)}.

Most weak measurement studies target the simplest non-trivial Hilbert space, of dimension two. Three-level or higher-dimensional discrete quantum systems have rarely been studied using the weak measurement formalism \cite{Dressel (2012),Dressel II (2012),Lorenzo (2015)}. A geometric representation of weak and modular values of their observables is lacking. However, recent applications of weak measurement theory in the context quantum computation research attest the interest of investigating weak values of high-level systems \cite{Xiao (2003),Xiao (2004)}. Weak values of qutrit observables show their usefulness in the experimental demonstration of the Kochen-Specker test of noncontextuality \cite{Jerger (2016)} and can be applied to the quantum Cheshire cat experiment \cite{Aharonov (2013)}. 

In this work, we explore the polar representation of weak and modular values in discrete quantum systems of arbitrary dimensions. We first express the modulus and the argument of weak and modular values of qubit observables in terms of vectors on the Bloch sphere to provide a purely geometric description of these values. For higher-dimensional $N$-level systems, we use the Majorana representation to describe their states by symmetric states of $N-1$ qubits \cite{Majorana (1932)}. Then, we proceed with demonstrating that an arbitrary weak or modular value of three-dimensional discrete quantum systems can be deduced from geometric quantities defined on the Bloch sphere. In particular, we find that both the modulus and the argument can be factored in two contributions, each connected to our results on qubit observables. Finally, we generalize our results to higher-dimensional systems. As an application of our new approach, we examine the phase discontinuities around singularities of weak values, which occur for orthogonal pre- and postselected states. We study the particular case of the weak value of a projector in a three-level system. In a second application, we  exploit the Majorana representation to revisit a well-known paradox previously studied by weak measurements: the quantum three-box paradox \cite{Aharonov (1991)}. We show that when the equivalent three-level system is recast as a pair of spin-$\frac{1}{2}$ particles in a symmetric spin state, this experiment involves contradictory conclusions about the entanglement state of the pre- and postselected particle pairs.  

\section{Polar representation of weak and modular values}

\subsection{Two-level quantum systems}

\begin{figure}[t]
	\begin{center}
			\includegraphics[width=1\textwidth]{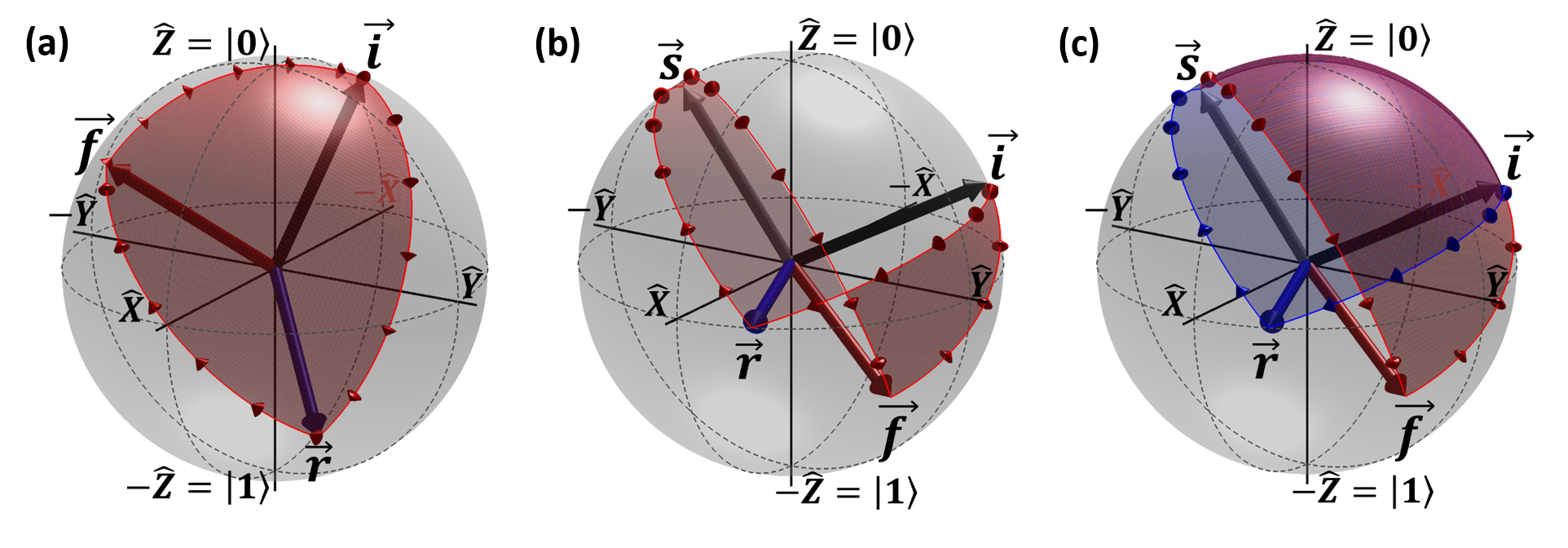} 
            \caption{Solid angle representations on the Bloch sphere. (a) The red solid angle $\Omega_{irf}$ of the sequence of states $|\phi_{i}\rangle\rightarrow|\phi_{r}\rangle\rightarrow|\phi_{f}\rangle\rightarrow|\phi_{i}\rangle$ is directly proportional to the argument of the weak value $\Pi_{r,w}$ (\ref{eq:PolarRepArg}). (b) The geometric component of the argument of the modular value $\sigma_{r,m}^{\alpha,\beta}$ is related to the red solid angle $\Omega_{i r s f}$. (c) The oriented spherical quadrangle that subtends  $\Omega_{i r s f}$ can be decomposed into two spherical triangles: the blue sequence $|\phi_{i}\rangle\rightarrow|\phi_{r}\rangle\rightarrow|\phi_{s}\rangle\rightarrow|\phi_{i}\rangle$ and the red sequence $|\phi_{i}\rangle\rightarrow|\phi_{s}\rangle\rightarrow|\phi_{f}\rangle\rightarrow|\phi_{i}\rangle$.} 
            \label{fig:bloch representation}
	\end{center}
\end{figure}

We start our developments by considering the two-level projection operator $\hat{\Pi}_{r}$ on the qubit state $|\phi_{r}\rangle$. This state is identified by the unit vector $\vec{r}\in{\rm I\!R}^{3}$ on the Bloch sphere. We consider an initial, preselected state $|\phi_{i}\rangle$ and a final, postselected state $|\phi_{f}\rangle$, defined by the unit vectors $\vec{i}$ and $\vec{f}$, respectively. The weak value  of the projector equals then
\begin{equation}\label{projectorweakvaluedef}
\Pi_{r,w} =\langle\phi_{f}| \phi_{r}\rangle \langle\phi_{r}|\phi_{i}\rangle \langle\phi_{f}|\phi_{i}\rangle^{-1}\:, 
\end{equation}
according to definition (\ref{eqWeakValueDef}). It can be uniquely expressed as a function of the three vectors defined on the Bloch sphere. Its modulus is given by
\begin{equation}\label{eq:PolarRepMod}
|\Pi_{r,w}|=\sqrt{\frac{1}{2}\frac{\left(1+\vec{f}\cdot\vec{r}\right)\left(1+\vec{r}\cdot\vec{i}\right)}{\left(1+\vec{f}\cdot\vec{i}\right)}}\:,
\end{equation}
and its argument by
\begin{equation}\label{eq:argprojwvalqubit}
\arg\Pi_{r,w}=\arctan\frac{\vec{f}\cdot\left(\vec{r}\times\vec{i}\right)}{1+\vec{f}\cdot\vec{r}+\vec{r}\cdot\vec{i}+\overrightarrow{f}\cdot\vec{i}} = -\frac{\Omega_{irf}}{2}\:,\label{eq:PolarRepArg}
\end{equation}
where $\Omega_{irf}$ is the oriented solid angle subtended at the center of the Bloch sphere by the geodesic triangle defined by the three vertices $\vec{i}$, $\vec{r}$ and $\vec{f}$ \cite{Eriksson (1990)}, as shown on figure \ref{fig:bloch representation}(a). The geodesic orientation is determined by the sequence of states $|\phi_{i}\rangle\rightarrow|\phi_{r}\rangle\rightarrow|\phi_{f}\rangle\rightarrow|\phi_{i}\rangle$. The complete expression of the weak value as a function of the three vectors on the Bloch sphere is derived in \ref{Modulus Calculation}.

The modulus is directly related to projection probabilities between the three qubit states. This is an obvious consequence of the definition of the weak value of a projector (\ref{projectorweakvaluedef}), which involves three bra-ket inner products. The modulus expression is quickly deduced from the correspondance linking the bra-ket inner product and the scalar product between Bloch vectors:
\begin{equation}\label{scalarproductcorrespondance}
\vert\langle \phi_{u}|\phi_{v}\rangle\vert=\sqrt{\frac{1}{2}(1+ \vec{u} \cdot \vec{v})}\: ,
\end{equation}
as shown in \ref{app:modulus expression}.

The argument of the weak value $\Pi_{r,w}$ (\ref{projectorweakvaluedef}) is equal to the quantum phase of the projection product $\langle \phi_{i}|\phi_{f}\rangle\langle \phi_{f}|\phi_{r}\rangle\langle \phi_{r}|\phi_{i}\rangle$, which is known as the three-vertex Bargmann invariant \cite{Bargmann (1964)}. Introduced by Bargmann for studying the difference between unitary and anti-unitary transformation, the quantity $\left(\prod_{k=1}^{N-1}\langle \phi_{k}|\phi_{k+1}\rangle\right)\langle \phi_{N}|\phi_{1}\rangle$ is invariant under gauge transformation and reparameterization. Mukunda and Simon showed by their kinematic approach \cite{Mukunda (1993),Mukunda (1993)bis} that the argument of the Bargmann invariant is related to the geometric phase $\gamma\left(C_{0}\right)$ \cite{Pancharatnam (1956),Berry (1984)} acquired in a closed loop $C_{0}$ on the Bloch sphere. The weak value is thus invariant under gauge transformations.

Now, we evaluate modular values in terms of vectors on the Bloch sphere. We consider an arbitrary observable of a two-level system and relate it to the Pauli and identity matrices by $\hat{A}=-\frac{1}{2} \beta\hat{I} +\frac{1}{2} \alpha \hat{\sigma}_{r}$, where $\alpha$, $\beta$ are real parameters and where by definition $\hat{\sigma}_{r}=\vec{r}\cdot\vec{\hat{\sigma}}=r_x \hat{\sigma}_x + r_y \hat{\sigma}_y + r_z \hat{\sigma}_z$. Physically, the normalized vector $\vec{r}$ on the Bloch sphere is equivalent to the direction of a spin measurement. An arbitrary modular value $\sigma_{r,m}^{\alpha,\beta}$ is then specified through the unitary operator: 
\begin{equation}
\hat{U}_{\sigma_{r}}^{\alpha,\beta}=e^{j\frac{\beta}{2}}e^{-j\frac{\alpha}{2}\hat{\sigma}_{r}}\,.
\end{equation}
The first component of the unitary operator applies a global $\frac{1}{2}\beta$-phase shift. The second component rotates all vectors on the Bloch sphere by an angle $\alpha$ around the $\vec{r}$-axis (in the Hilbert state space representation of the qubit, the rotation angle is $\frac{1}{2}\alpha$). Its modular value  is thus equal to
\begin{equation}\label{2lvlmodularvaluedef}
\sigma_{r,m}^{\alpha,\beta}=e^{j\frac{\beta}{2}} \langle\phi_{f}|e^{-j\frac{\alpha}{2}\hat{\sigma}_{r}} |\phi_{i}\rangle \langle\phi_{f}|\phi_{i}\rangle^{-1}\:, 
\end{equation}
according to definition (\ref{eqModularValueDef}). We define the state $|\phi_{s}\rangle=e^{-j\frac{\alpha}{2}\hat{\sigma}_{r}}|\phi_{i}\rangle$, which result from applying the rotation operator to the initial, preselected state. After rotating the initial vector $\vec{i}$ around the $\vec{r}$-axis by the angle $\alpha$, we obtain the vector $\vec{s}$ characterizing $|\phi_{s}\rangle$ on the Bloch sphere. It is given by Rodrigue's rotation formula:
\begin{equation}\label{eq:RodrigueRotFormula}
\vec{s}=\cos\alpha\ \vec{i} + \vec{r}\cdot\vec{i}\ (1 -\cos\alpha)\ \vec{r}+\sin\alpha\ \vec{r}\times\vec{i}\,.
\end{equation}
We deduce immediately the modulus of the modular value $\vert \sigma_{r,m}^{\alpha,\beta} \vert=\vert \langle\phi_{f}|\phi_{s}\rangle \langle\phi_{f}|\phi_{i}\rangle^{-1}\vert$ from the correspondance (\ref{scalarproductcorrespondance}) linking the inner product in Hilbert space to the scalar product between Bloch vectors:
\begin{equation}
|\sigma_{r,m}^{\alpha,\beta}|= \sqrt{\frac{1+\vec{f}\cdot\vec{s}}{1+\vec{f}\cdot\vec{i}}}\:.
\end{equation}

During its rotation around the $\vec{r}$-axis, the trajectory of the initial vector $\vec{i}$ follows a non-geodesic open arc on the Bloch sphere \cite{Martinez (2012)}, contrary to the projector case (which involved solely geodesic arcs). Consequently, the rotated state $|\phi_{s}\rangle$ is no longer in phase with the initial state $|\phi_{i}\rangle$. The argument of the modular value is therefore evaluated either by following the reasoning detailed in \cite{Cormann (2015)}, or by finding the complete expression of the modular value as a function of Bloch vectors (see \ref{Qubit unitary operator}): 
\begin{equation}
\arg \sigma_{r,m}^{\alpha,\beta} =\frac{\beta-\alpha}{2}-\frac{\Omega_{irsf}}{2}\:.\label{eq:MVPhase}
\end{equation}
The argument contains a dynamical contribution that depends on the parameters $\alpha$ and $\beta$, and a geometric phase $-\frac{1}{2}\Omega_{irsf}$ that depends solely on vectors defined on the Bloch sphere. The exact expression of the geometric phase as a function of Bloch vectors is given by (\ref{eq:gmphasemodularirsf})  in \ref{Qubit unitary operator}. The dynamical contribution can vanish by choosing $\beta$ equal to $\alpha$. The geometric phase $\Omega_{i r s f}$  depends on the oriented spherical quadrangle delimited by the four vectors $\vec{i}$, $\vec{r}$, $\vec{s}$ and $\vec{f}$ on the Bloch sphere, as depicted on  figure \ref{fig:bloch representation}(b). The orientation is defined by the corresponding sequence of states $|\phi_{i}\rangle\rightarrow|\phi_{r}\rangle\rightarrow|\phi_{s}\rangle\rightarrow|\phi_{f}\rangle\rightarrow|\phi_{i}\rangle$. As illustrated on figure \ref{fig:bloch representation}(c), the oriented spherical quadrangle ${i \rightarrow r \rightarrow s \rightarrow f  \rightarrow i}$ can be written using the two oriented spherical triangles ${i \rightarrow r \rightarrow s  \rightarrow i}$ (blue curve) and ${i \rightarrow s \rightarrow f  \rightarrow i}$ (red curve), as the paths $i\rightarrow s$ and $s\rightarrow i$ present in these triangles cancel each other. The solid angles that they subtend at the center of the Bloch sphere are thus related by:
\begin{equation}\label{eq:quadto2trig}
\Omega_{i r s f}=\Omega_{irs}+\Omega_{isf}\:.
\end{equation}
The sequence of states associated to the solid angle $\Omega_{irs}$ arises from the application of the operator $e^{-j\frac{\alpha}{2}\hat{\sigma}_{r}}$, while the solid angle $\Omega_{isf}$ is associated to the postselection of $|\phi_{f}\rangle$. Thanks to this decomposition, the argument of the modular value can be evaluated using expression (\ref{eq:argprojwvalqubit}), found for the argument of the projector weak value. This property will prove to be useful for higher-level quantum systems. Refer to \ref{Qubit unitary operator} for more details and proof of (\ref{eq:quadto2trig}).
 
To close this discussion of two-level systems, we note that the modulus and the argument of the weak value of an arbitrary spin observable  $\hat{\sigma}_{r}$ can be obtained from the modular value by setting $\alpha=\beta=\pi$ because then $A_w=A_m$ \cite{Cormann (2015)}. 

\subsection{Three- and higher-level quantum systems}

In the previous section, we derived geometric expressions for weak and modular values of qubit observables. This was made possible thanks to the unique one-to-one correspondance linking the qubit states in two-dimensional Hilbert space and the vectors on the unit sphere in three-dimensional physical space. Unfortunately, such a correspondance does not exist for higher-level systems: their states cannot be identified bijectively with the vectors on a unit sphere in a higher-dimensional real vectorial space. However, following an approach developed by Majorana \cite{Majorana (1932)}, it is possible to represent states of a $N$-level system by $N-1$ vectors on the Bloch sphere. With this essential insight, we will now be able to find geometric expressions for weak and modular values of observables of three-level quantum systems, which can be easily generalized to arbitrary $N$-level systems. 

According to the Majorana approach, amongst the pure quantum states of a system of $N-1$ qubits, it is possible to distinguish a class of states which are symmetric with respect to all possible permutations of the $N-1$ qubit subsystems. This class of symmetric pure quantum states can be identified with the set of all states of a single system described in a $N$-dimensional Hilbert space \cite{Ganczarek (2012), Bloch (1945)}. An arbitrary symmetric state of this set $|\Psi\rangle$ can be written as:
\begin{equation}
|\Psi\rangle=K \sum_{P}\hat{P}\left[|\phi^{(1)}\rangle|\phi^{(2)}\rangle...|\phi^{(N-1)}\rangle\right]\:,\label{eq:MajoranaRepresentation}
\end{equation}
where $|\phi^{(k)}\rangle$, with $k=1,2,...,N-1$, denotes the $k^{th}$ qubit state, $\sum_P \hat{P}$ corresponds to the set of all $(N-1)!$ permutations of the qubits and $K$ is the normalization factor. The state $|\Psi\rangle$ is determined by an unordered set of $N-1$ points on the Bloch sphere, called the Majorana points.

Tamate \textit{et al.} \cite{Tamate (2011)} demonstrated that a set of three symmetric states $|\Psi_{1}\rangle$, $|\Psi_{2}\rangle$ and $|\Psi_{3}\rangle$ of a ensemble of $N-1$ qubits can always be transformed by an appropriate unitary transformation to the following specific set of symmetric states:
\begin{eqnarray}
|\Psi_{1}''\rangle=K \sum_{P}\hat{P}\left[|\phi^{(1)}_{1}\rangle...|\phi^{(N-1)}_{1}\rangle\right]\:,\nonumber \\
|\Psi_{2}''\rangle=\underbrace{|\phi_{2}\rangle...|\phi_{2}\rangle}_{N-1}\:,\:\:\:\:\:\:\:\: 
|\Psi_{3}''\rangle=\underbrace{|\phi_{3}\rangle...|\phi_{3}\rangle}_{N-1}\:.
\label{eq:AlgStr}
\end{eqnarray}
After this unitary transformation, the states $|\Psi_{2}''\rangle$ and $|\Psi_{3}''\rangle$ are factored in products of $N-1$ identical qubit states. Only $|\Psi_{1}''\rangle$ remains in an entangled state of $N-1$ qubits. Thus,  $|\Psi_{1}''\rangle$ is represented by $N-1$, generally distinct, points $\vec{p}_1^{\: (k)}$ on the Bloch sphere, while the states $|\Psi_{2}''\rangle$ and $|\Psi_{3}''\rangle$ are described by single degenerate points, $\vec{p}_2$ and $\vec{p}_3$, respectively. Consequently, the argument of the corresponding three-vertex Bargmann invariant $\langle\Psi_{1}|\Psi_{2}\rangle \langle\Psi_{2}|\Psi_{3}\rangle \langle\Psi_{3}|\Psi_{1}\rangle$ is expressed by a sum of $N-1$ geometric phases \cite{Tamate (2011)}:
\begin{equation}
\gamma\left(\Psi_{1},\Psi_{2},\Psi_{3}\right)=\sum_{k=1}^{N-1} \gamma(\phi_{1}^{(k)},\phi_{2},\phi_{3})\:,\label{eq:DSGP}
\end{equation}
where each term of this sum is equal to the solid angle $- \frac{1}{2}\Omega_{123}^{(k)}$ defined by the corresponding three vectors $\vec{p}_1^{\: (k)}$, $\vec{p}_2$ and $\vec{p}_3$ through relation (\ref{eq:PolarRepArg}).

\subsubsection{Weak values of projectors in qutrit systems}

With this knowledge, we proceed now with the evaluation of the weak value $\Pi_{r,w}^{(3)}$ of the projector on an arbitrary state $|\psi_{r}\rangle$ of a three-level quantum system. Similarly to the two-level case, the weak value $\Pi_{r,w}^{(3)}$ of a three-level quantum system involves a set of three qutrit states $|\psi_{i}\rangle$, $|\psi_{r}\rangle$ and $|\psi_{f}\rangle$. Their Majorana representation in terms of symmetric two-qubit states are given by $|\Psi_{i}\rangle$, $|\Psi_{r}\rangle$ and $|\Psi_{f}\rangle$, repectively. Through a unitary transformation $U$, we transform these states to the set 
\begin{eqnarray}\label{eq:AlgStr}
|\Psi_{i}''\rangle=K_i \left[|\phi^{(1)}_{i}\rangle |\phi^{(2)}_{i}\rangle + |\phi^{(2)}_{i}\rangle |\phi^{(1)}_{i}\rangle\right]\:, \\
|\Psi_{f}''\rangle=|\phi_{f}\rangle |\phi_{f}\rangle \:,\:\:\:\:\:\:\:\: 
|\Psi_{r}''\rangle=|\phi_{r}\rangle |\phi_{r}\rangle\:. \nonumber
\end{eqnarray}
where the normalization factor $K_{i}=(2 + 2\:   |\langle\phi^{(2)}_{i}|\phi^{(1)}_{i}\rangle|^2 )^{-\frac{1}{2}}$. The form taken by the unitary transform and the exact expression of the different qubit states will be determined quantitatively later. Indeed, their formulation is not needed to obtain the researched expression as a function of vectors on the Bloch sphere. We can now evaluate the weak value, which is invariant under the unitary transformation:
\begin{equation}
\Pi_{r,w}^{(3)}=\frac{\langle\Psi_{f}''| \Psi_{r}''\rangle \langle\Psi_{r}''|\Psi_{i}''\rangle} {\langle\Psi_{f}''|\Psi_{i}''\rangle}= \frac{\langle\phi_{f}| \phi_{r}\rangle^2 \langle\phi_{r}|\phi_{i}^{(1)}\rangle \langle\phi_{r}|\phi_{i}^{(2)}\rangle} {\langle\phi_{f}|\phi_{i}^{(1)}\rangle \langle\phi_{f}|\phi_{i}^{(2)}\rangle}\:, 
\end{equation}
In the Majorana representation, the weak value of a qutrit projector is thus given by the product of two weak values of a qubit projector (\ref{projectorweakvaluedef}), but for different initial states $|\phi_i^{(1)}\rangle$ and $|\phi_i^{(2)}\rangle$. Using the expressions obtained for the modulus (\ref{eq:PolarRepMod}) and the argument (\ref{eq:PolarRepArg}) of weak values of qubit projectors, we obtain immediately the modulus the weak value of the qutrit projector:
\begin{equation}
|\Pi_{r,w}^{(3)}|=\sqrt{\frac{1}{2}\frac{\left(1+\vec{f}\cdot\vec{r}\right)\left(1+\vec{r}\cdot\vec{i}_{2}\right)}{\left(1+\vec{f}\cdot\vec{i}_{2}\right)}}\sqrt{\frac{1}{2}\frac{\left(1+\vec{f}\cdot\vec{r}\right)\left(1+\vec{r}\cdot\vec{i}_{1}\right)}{\left(1+\vec{f}\cdot\vec{i}_{1}\right)}}\:,
\end{equation}
as well as its argument:
\begin{equation}
\arg\Pi_{r,w}^{(3)}=-\frac{\Omega_{i_{2}rf}}{2}-\frac{\Omega_{i_{1}rf}}{2}\:,
\end{equation}
where the four relevant qubit states were described with their vectors on the Bloch sphere in an obvious notation. Interestingly, the three-level weak value is determined by two independent sequences of the Bloch vectors $\vec{i}_{1}$ and $\vec{i}_{2}$. The modulus $\Pi_{r,w}^{(3)}$ is given by the product of two square roots. Each ratio inside a square root represents the projection probability that the initial vectors $\vec{i}_{k}$ aligns with the final Bloch vector $\vec{f}$ by passing through the intermediate vector $\vec{r}$, divided by the projection probability that the initial vectors $\vec{i}_{k}$ aligns directly with the final Bloch vector $\vec{f}$ . The argument is proportional to the sum of two solid angles $\Omega_{i_{1}rf}$ and $\Omega_{i_{2}rf}$ delimited by the geodesic triangles on the Bloch sphere with the three vertices $\vec{i}_{1}$, $\vec{r}$, $\vec{f}$ and $\vec{i}_{2}$, $\vec{r}$, $\vec{f}$, respectively. 

We now construct the unitary transformation $U$ in order to determine the qubit states. The normalized state $|\psi_{r}\rangle$ is written as a function of four real parameters $\theta$, $\epsilon$, $\chi_{1}$ and $\chi_{2}$ so that $|\psi_{r}\rangle=\left(e^{j\chi_{1}}\:\cos\epsilon\,\sin\theta,\, e^{j\chi_{2}}\: \sin\epsilon\,\sin\theta,\, \cos\theta\right)^{T}$. We define the unitary operator $\hat{U}^{(1)}\in U(3)$ which maps $|\psi_{r}\rangle$ to the state $|\psi_{r}'\rangle =\left(0,\, 0,\, 1\right)^{T}$:
\begin{equation}
\hat{U}^{(1)}=\left( \begin{array}{ccc}
-e^{-j\chi_{1}}\sin\epsilon & e^{-j\chi_{2}}\cos\epsilon & 0 \\
-e^{-j\chi_{1}}\cos\epsilon\cos\theta & -e^{-j\chi_{2}}\sin\epsilon\cos\theta & \sin\theta \\
e^{-j\chi_{1}}\cos\epsilon\sin\theta & e^{-j\chi_{2}}\sin\epsilon\sin\theta & \cos\theta \end{array} \right)\:.
\end{equation}
It also induces the transformations $|\psi_{i}\rangle \rightarrow |\psi_{i}'\rangle$ and $|\psi_{f}\rangle \rightarrow |\psi_{f}'\rangle$. As we shall see later, the resulting state 
$|\psi_{r}'\rangle$ is associated to the factored state $|\Psi_{r}'\rangle=|0\rangle|0\rangle$ in the Majorana representation. This state presents two overlapping Majorana points on the Bloch sphere's north pole. We consider now a second unitary operator $\hat{U}^{(2)}\in U(3)$ which leaves $|\psi_{r}'\rangle$ invariant, but transforms the postselected state $|\psi_{f}'\rangle$ into a separable two-qubit state in the Majorana representation. In particular, we rewrite $|\psi_{f}'\rangle$ using a general expression
\footnote{The unitary operator $\hat{U}^{(1)}$ generally adds a phase factor to all three components of the state vector. However, as the weak value is gauge invariant, we can remove arbitrarily the phase factor from the third component without loss of generality. Note that this operation does not preserve the phase of the inner product beween two states. Therefore the unitary operator must be applied to all states involved in the weak value expression, and all global phases sould be removed accordingly.}
with the four real parameters  $\eta$, $\delta$, $\xi_{1}$ and $\xi_{2}$ so that $|\psi_{f}'\rangle=\left(e^{j\xi_{1}}\:\cos\delta\,\sin\eta,\, e^{j\xi_{2}}\: \sin\delta\,\sin\eta,\, \cos\eta\right)^{T}$. This unitary transformation is given by:
\begin{equation}\label{eq:U2unitopdef}
\hat{U}^{(2)}=\left( \begin{array}{ccc}
e^{-j\xi_{1}}\cos\alpha & e^{-j\xi_{2}}\sin\alpha & 0 \\
e^{-j\xi_{1}}\sin\alpha & -e^{-j\xi_{2}}\cos\alpha & 0 \\
0 & 0 & 1 \end{array} \right)\:,
\end{equation} 
with $\alpha=\delta + \arccos(\tan\frac{\eta}{2})$. After this unitary transformation, the postselected state becomes $|\psi_{f}''\rangle=(1-\cos\eta,\, \sqrt{2\cos\eta(1-\cos\eta)},\, \cos\eta)^{T}$. As will be explained later, its Majorana representation is given by the factored state $|\Psi_{f}''\rangle=|\phi_{f}\rangle|\phi_{f}\rangle$, where $|\phi_{f}\rangle=\sqrt{\cos\eta}|0\rangle + \sqrt{1-\cos\eta}|1\rangle$. Due to the sequential application of the transformations $\hat{U}^{(1)}$ and $\hat{U}^{(2)}$, the initial three-level state $|\psi_{i}\rangle$ evolves to the normalized state $|\psi_{i}''\rangle=c_{0} |0\rangle + c_{1} |1\rangle + c_{2} |2\rangle$. Its Majorana representation can be obtained by solving the Majorana polynomial \cite{Bloch (1945),Usha (2012)}: 
\begin{equation}
\frac{c_0}{\sqrt{2}}-c_1 z +\frac{c_2}{\sqrt{2}}\, z^2=0\:.
\end{equation}
The two roots $z_k$ of this polynomial are related to the polar and azimutal angle on the Bloch sphere by
\begin{equation}\label{eq:majpolroots}
z_k=e^{j \phi_k} \tan \frac{\theta_k}{2}  \:.
\end{equation}
Separable states occur when the disciminant of the polynomial is nul, so that the roots are identical. The Majorana representation associates the following states together: $|0\rangle \rightarrow |\Psi_0\rangle = |1\rangle|1\rangle$,  $|2\rangle \rightarrow |\Psi_2\rangle = |0\rangle|0\rangle$ and  $|1\rangle \rightarrow |\Psi_1\rangle = 2^{-\frac{1}{2}}(|0\rangle|1\rangle+|1\rangle|0\rangle)$. In the end, we find thus that the projector state was mapped to the Bloch sphere vector  $\vec{r}=(0,0,1)$, the postselected, final state was mapped to the vector $\vec{f}=(\sqrt{4\cos\eta(1-\cos\eta)},0,2\cos\eta-1)$. The initial state is given by $|\Psi_{i}''\rangle=K_{i}\left[|\phi^{(1)}_{i}\rangle|\phi^{(2)}_{i}\rangle +|\phi^{(2)}_{i}\rangle|\phi^{(1)}_{i}\rangle\right]$ where the two qubits states are deduced from the roots (\ref{eq:majpolroots}) of the Majorana polynomial $|\phi^{(k)}_{i}\rangle=\cos \frac{\theta_k}{2}  |0\rangle +  \sin \frac{\theta_k}{2} e^{j \phi_k}|1\rangle$ and where the normalization factor can also be evaluated to $K_i=1/\sqrt{3+ \vec{i}_1 \cdot \vec{i}_2}$.
 

\subsubsection{Modular values in qutrit systems}

The same kind of relations can also be established for the modular value of an arbitrary three-level evolution operator:
\begin{equation}
\hat{U}_{\lambda_{r}}^{\alpha,\beta}=e^{j \beta}e^{-j \alpha \hat{\lambda}_{r}}\,,
\end{equation}
where $\hat{\lambda}_{r}=\vec{r}_{(8)}\cdot\vec{\hat{\lambda}}$ with $\vec{r}_{(8)}\in{\rm I\!R}^{8}$ a normalized vector pointing in a $8$-dimensional space. The $k^{th}$ element ($k=1,2,...,8$) of the vector $\vec{\hat{\lambda}}$ corresponds to the Gell-Mann operator $\hat{\lambda}_k$. A summary containing the essential properties of the Gell-Mann operators is presented in reference \cite{Goyal (2016)}. For our purposes, it suffices to know that the hermitian operator $\hat{\lambda}_{r}$ is traceless, that the trace of  $\hat{\lambda}_{r}^2$ equals 2, and that, when it verifies the condition $\det(\hat{\lambda}_{r})=0$, its eigenvalues are $-1$, $0$ and $+1$. The parameter $\beta$ induces a phase shift while the parameter $\alpha$ was defined so that it corresponds to the rotation angle when $\hat{\lambda}_{r}$ is a spin-1 operator.
 
 We consider the set of three qutrits states $|\psi_{i}\rangle$, $|\psi_{r}\rangle$ and $|\psi_{f}\rangle$, where $|\psi_{i}\rangle$ is the initial, preselected state, $|\psi_{f}\rangle$ is the final, postselected state and $|\psi_{r}\rangle$ is an eigenvector state of the operator $\hat{\lambda}_r$, associated with an eigenvalue $\lambda_r$. Any eigenvector can be selected but we could arbitrarily select the largest eigenvalue to remain in line with the spirit of the developments we followed for the qubit case. The modular value $\lambda_{r,m}^{\alpha,\beta}$ of the Gell-Mann operator defined through the previous unitary operator becomes
 \begin{equation}\label{3lvlmodularvaluedef}
\lambda_{r,m}^{\alpha,\beta}=e^{j \beta} \langle\psi_{f}|e^{-j \alpha\hat{\lambda}_{r}} |\psi_{i}\rangle \langle\psi_{f}|\psi_{i}\rangle^{-1}\: ,
\end{equation}
 according to definition (\ref{eqModularValueDef}). Following the procedure developed for two-level systems, we define the state $|\psi_{s}\rangle=e^{-j \alpha \hat{\lambda}_{r}}|\psi_{i}\rangle$, which result from applying the $\alpha$-evolution operator to the initial, preselected state. As for the qutrit projector case, there exists a couple of unitary operators $\hat{U}^{(1)}, \hat{U}^{(2)}\in U(3)$ transforming the eigenvector state $|\psi_{r}\rangle$ and the postselected state $|\psi_{f}\rangle$ to $|\Psi_{r}''\rangle=|0\rangle|0\rangle$ and $|\Psi_{f}''\rangle=|\phi_{f}\rangle|\phi_{f}\rangle$, respectively. Additionally, the initial state $|\psi_{i}\rangle$ is mapped to the state $|\Psi_{i}''\rangle=K_{i}\left[|\phi^{(1)}_{i}\rangle|\phi^{(2)}_{i}\rangle +|\phi^{(2)}_{i}\rangle|\phi^{(1)}_{i}\rangle\right]$ while the state $|\psi_{s}\rangle$ is associated to the state $|\Psi_{s}''\rangle=K_{s}\left[|\phi^{(1)}_{s}\rangle|\phi^{(2)}_{s}\rangle +|\phi^{(2)}_{s}\rangle|\phi^{(1)}_{s}\rangle\right]$.
Therefore, the modular value is expressed by
\begin{equation}\label{eq:3lvlmodularvaluedefMajorana}
\lambda_{r,m}^{\alpha,\beta}=e^{j \beta} \frac{K_s}{K_i} \frac{\langle\Psi_{f}''|\Psi_{s}''\rangle}{ \langle\Psi_{f}''|\Psi_{i}''\rangle}=e^{j \beta} \frac{K_s}{K_i} \frac{\langle\phi_{f}|\phi_{s}^{(1)}\rangle \langle\phi_{f}|\phi_{s}^{(2)}\rangle}{ \langle\phi_{f}|\phi_{i}^{(1)}\rangle \langle\phi_{f}|\phi_{i}^{(2)}\rangle}\: ,
\end{equation}
which contains the contributions of two modular values of qubits. This factorisation is very similar to the one obtained in the qutrit projector case. Consequently, the modulus of the modular value $\lambda_{r,m}^{\alpha,\beta}$ is given as a function of vectors on the Bloch sphere according to the expression
\begin{equation}
|\lambda_{r,m}^{\alpha,\beta}|=\frac{K_{s}}{K_{i}}\sqrt{\frac{\left(1+\vec{f}\cdot\vec{s}_{2}\right)}{\left(1+\vec{f}\cdot\vec{i}_{2}\right)}}\sqrt{\frac{\left(1+\vec{f}\cdot\vec{s}_{1}\right)}{\left(1+\vec{f}\cdot\vec{i}_{1}\right)}}\:,
\end{equation}
where $K_{n}=1/\sqrt{3+\vec{n}_{2}\cdot\vec{n}_{1}}$ (with $n=i,s$), while its argument is found to be
\begin{equation}\label{eq:argmodvalqutrit}
\arg\lambda_{r,m}^{\alpha,\beta}=\beta- \alpha  \lambda_r -\frac{\Omega_{i_{2}rs_{2}f}}{2}-\frac{\Omega_{i_{1}rs_{1}f}}{2}\:, 
\end{equation}
where the solid angles were defined similarly to the qubit case. The vectors $\vec{s}_{1}$ and $\vec{s}_{2}$ and $\vec{i}_{1}$ and $\vec{i}_{2}$ can be found by solving the Majorana polynomial for the states $|\psi_{s}''\rangle$ and $|\psi_{i}''\rangle$, respectively. More details about the procedure leading to (\ref{eq:argmodvalqutrit}) can be found in \ref{App:argModValqutrit}.

Because the algebraic structure of the Gell-Mann $\hat{\lambda}$-operators is significantly different from the structure of the Pauli $\hat{\sigma}$-operators, the weak value of the $\hat{\lambda}_r$ observable cannot be evaluated from its modular value simply by setting particular values for $\alpha$ and $\beta$, contrary to what was possible in the qubit case. It is however possible to express the modular value as a function of weak values of $\hat{\lambda}_r$ and $\hat{\lambda}_r^2$ in a closed form. For example, in the simple case of a spin-1 observable (corresponding to $\det(\hat{\lambda}_{r})=0$), setting the phase shift $\beta$ to zero, the relationship between weak and modular values of  $\hat{\lambda}_r$ is deduced readily from the exact value of the exponential operator: $e^{-j\alpha\hat{\lambda}_r}=1-j \sin \alpha\: \hat{\lambda}_r +  (\cos \alpha -1)\:\hat{\lambda}_r^2$ \cite{Curtright (2015)}, which can be obtained using the Cayleigh-Hamilton theorem.

\subsubsection{Generalization to arbitrary $N$-level systems}

As for the three-level quantum system, any set of three $N$-level quantum states can be transformed to the specific set (\ref{eq:AlgStr}) 
\begin{eqnarray}
|\Psi_{i}''\rangle=K_i \sum_{P}\hat{P}\left[|\phi^{(1)}_{i}\rangle...|\phi^{(N-1)}_{i}\rangle\right]\:,\nonumber \\
|\Psi_{f}''\rangle=\underbrace{|\phi_{f}\rangle...|\phi_{f}\rangle}_{N-1}\:,\:\:\:\:\:\:\:\: 
|\Psi_{r}''\rangle=\underbrace{|\phi_{r}\rangle...|\phi_{r}\rangle}_{N-1}\:,
\label{eq:AlgStrgeneralized}
\end{eqnarray}
by applying the appropriate unitary transformations $\hat{U}^{(1)}$, $\hat{U}^{(2)}\in U(N)$. Consequently, weak values of a $N$-level pre- and postselected projector are always deduced by the product of $N-1$ square roots of a probability ratio for the modulus and the sum of $N-1$ spherical triangles for the argument, by introducing $N-1$ initial two-level states:
\begin{eqnarray}
|\Pi_{r,w}^{(N)}|=|\Pi_{1,w}|\cdot|\Pi_{2,w}|\cdot ...\cdot|\Pi_{N-1,w}|\:,\label{eq:GeneralModulus}\\ 
\arg\Pi_{r,w}^{(N)}=\arg\Pi_{1,w}+\arg\Pi_{2,w}+ ... + \arg\Pi_{N-1,w}\:.\label{eq:GeneralSolidAngle}
\end{eqnarray}  
For modular values, this generalization remains valid if they are defined using traceless Hermitian operators $\hat{\Lambda}_r$. The associated unitary operator is $\hat{U}_{\Lambda_{r}}^{\alpha,\beta}=e^{j \beta}e^{-j \alpha \frac{N-1}{2} \hat{\Lambda}_{r}}$. The Majorana representation of the state $|\psi_{s}\rangle=\hat{U}_{\Lambda_{r}}^{\alpha,\beta}|\psi_{i}\rangle$ introduces the additional set of $N-1$ Bloch vectors $\vec{s}_{k}$ (with $k=1,...,N-1$), so that  
\begin{eqnarray}
|\Lambda_{r,m}^{\alpha,\beta}|=\frac{K_{s}}{K_{i}}\sqrt{\prod_{k=1}^{N-1}\frac{\left(1+\vec{f}\cdot\vec{s}_{k}\right)}{\left(1+\vec{f}\cdot\vec{i}_{k}\right)}} \:,\\
\arg\Lambda_{r,m}^{\alpha,\beta}=\beta-\alpha  \frac{N-1}{2}  \Lambda_r -\sum_{k=1}^{N-1}\frac{\Omega_{i_{k}rs_{k}f}}{2}\:, 
\end{eqnarray}
where the parameter $\alpha$ is defined to respect convention on angular momentum for spin operators and where $\vert\psi_r\rangle$ is an arbitrary eigenvector of $\hat{\Lambda}_r$ with eigenvalue $\Lambda_r$.

\section{Applications involving three-level quantum systems}

\subsection{Singularities in weak values}

Here, we examine the discontinuous behavior around singularities of weak values of three-level projectors. They occur when the preselected and postselected states are orthogonal to each other, as the denominator of the weak value (\ref{eqWeakValueDef}) diverges then. We will show that  this discontinuity is caused by the geometric phase.

For this purpose, we fix three-level projector to $|\psi_{r}\rangle=(0,0,1)^{T}$ and pick the particular final state $|\psi_{f}\rangle=\frac{1}{2}(\sqrt{2}\,,1\,,1)^{T}$. An arbitrary initial state can then be written in the form of $|\psi_{i}\rangle=(e^{j\chi_{2}}\sin\epsilon\sin\theta,e^{j\chi_{1}}\cos\epsilon\sin\theta,\cos\theta)^{T}$. The projector weak value is given by $\Pi_{r,w}^{(3)}=[1+\tan\theta\,(\sqrt{2} \sin\epsilon\, e^{j\chi_{2}}+\cos\epsilon\, e^{j\chi_{1}})]^{-1}$. The set of three states  $|\psi_{i}\rangle$, $|\psi_{r}\rangle$ and $|\psi_{f}\rangle$ is transformed to the specific set (\ref{eq:AlgStr}) by applying the unitary operator:  
\begin{equation}
\hat{U}^{(2)}=\left( \begin{array}{ccc}
0 & 1 & 0 \\
1 & 0 & 0 \\
0 & 0 & 1 \end{array} \right)\:.
\end{equation}
This operator corresponds to the $\hat{U}^{(2)}$ operator defined in (\ref{eq:U2unitopdef}) with its parameters set to $\alpha=\frac{\pi}{2}$ and $\xi_{1}=\xi_{2}=0$. In this way, the final state $|\Psi_{f}''\rangle$ becomes $|x_{+}\rangle|x_{+}\rangle$ in the Majorana representation, with $|x_{+}\rangle=\frac{1}{\sqrt{2}}(|0\rangle+|1\rangle)$, while the projector state $|\Psi_{r}''\rangle$= $|0\rangle|0\rangle$. With these states, the projector is represented by the vector $\vec{e}_z$ on the Bloch sphere, while the final state is represented by the vector $\vec{e}_x$. The initial state evolves to $|\psi_{i}''\rangle=(e^{j\chi_{1}}\cos\epsilon\sin\theta,e^{j\chi_{2}}\sin\epsilon\sin\theta,\cos\theta)^{T}$.

To find the Majorana representation $|\Psi_{i}''\rangle$ of the initial state, we need to solve its Majorana polynomial. The solutions are cumbersome for an arbitrary initial state (see \ref{appendix:singularities}). To gain physical insight, we adopt here a simplified set of parameters to describe the initial state: $\epsilon=\arcsin (\tan \frac{\pi}{6})$ and $\chi_1=2 \chi_2=\frac{4}{3}\pi$. With these parameters, the initial state becomes $|\psi_{i}''\rangle=(e^{-j\frac{2}{3}\pi} \sqrt{\frac{1}{3}} \sin\theta, e^{j\frac{2}{3}\pi} \sqrt{\frac{2}{3}} \sin\theta, \cos\theta)^{T}$ and the weak value is a real number that depends only on the parameter $\theta$ so that $\Pi_{r,w}^{(3)}=(1-\sqrt{\frac{2}{3}}\tan\theta)^{-1}$. The roots of the corresponding second-degree Majorana polynomial are
\begin{equation}\label{eq:app1roots}
z_{1,2}=\tan(\frac{\beta_{1,2}}{2})\,e^{j\alpha_{1,2}}=e^{-j\frac{1}{3} \pi } \frac{-\tan \theta \pm \sqrt{ (\tan \theta -2 \sqrt{6} ) \tan \theta}}{\sqrt{6}}\: .
\end{equation}
The two qubits states are thus given by $|\phi_{i}^{(1,2)}\rangle=\cos\left(\frac{\beta_{1,2}}{2}\right) |0\rangle + \sin\left(\frac{\beta_{1,2}}{2}\right)e^{j\alpha_{1,2}} |1\rangle$, with the corresponding Bloch vectors $\vec{i}_{1,2}=(\cos\alpha_{1,2}\sin\beta_{1,2} ,\sin\alpha_{1,2} \sin\beta_{1,2},\cos\beta_{1,2})$, where $\alpha_{1}$, $\alpha_{2}$  are the azimuth angles and $\beta_{1}$, $\beta_{2}$ the polar angles.

\begin{figure}[t!]
	\begin{center}
			\includegraphics[width=0.7\textwidth]{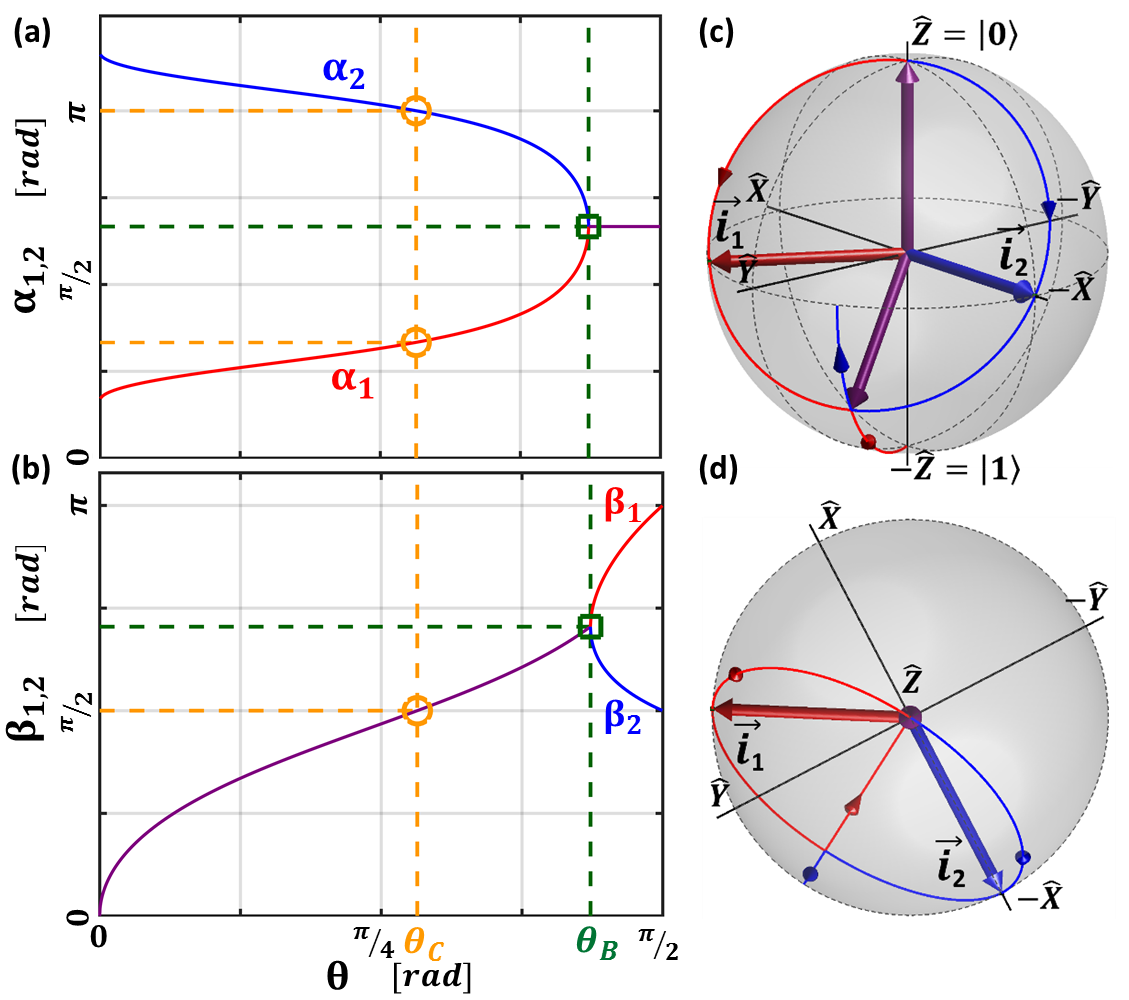} 
            \caption{(a-b) Evolution of the angles $\alpha_{1,2}$ and $\beta_{1,2}$ characterizing the initial qubit states $|\phi_{i}^{(1,2)}\rangle$ with respect to $\theta$ for $\epsilon\approx 0.19(6)\pi$, $\chi_{1}=\frac{4\pi}{3}$ and $\chi_{2}=\frac{2\pi}{3}$. The bifurcation is represented by the green square: (a) for $\theta>\theta_B$, the values of the azimuth angles $\alpha_{1}$ (red) and $\alpha_{2}$ (blue) are degenerate (violet), (b) while for $\theta<\theta_B$, the polar angles $\beta_{1}$ (red) and $\beta_{2}$ (blue) are degenerate (violet). (c-d) Representation of the corresponding trajectories of the vectors $\vec{i}_{1,2}$ on the Bloch sphere in front (c) and bird's-eye views (d). The orientation of the illustrated red and blue Bloch vector corresponds to the particular value $\theta=\theta_C$, for which the weak value diverges.} 
            \label{fig:Protocol}
	\end{center}
\end{figure}\noindent

In figure \ref{fig:Protocol}(a-b), we observe a bifurcation in the values of $\alpha_{1,2}$ and $\beta_{1,2}$, when expressed as a function of $\theta$ (green squares). This bifurcation occurs when the discriminant of the Majorana polynomial equals zero, for $\tan\theta_B=2 \sqrt{6}$ so that $\theta_{B}\approx 0.43(6)\pi$. At this particular value, the initial state $|\psi_{i}''\rangle$ is a product state of two identical qubits. For all parameters $\theta<\theta_B$, the polar angles $\beta_{1}$ and $\beta_{2}$ are degenerate (violet). In contrast, the values of the azimuth angles $\alpha_{1}$ (red) and $\alpha_{2}$ (blue) are initially different, but then symmetrically reach the joint value $\chi_{1}/2$ (violet). This behavior result from the square root of the discriminant in (\ref{eq:app1roots}) being a pure imaginary number for $\theta<\theta_B$: the two roots pick up opposite phase with respect to the global phase factor while their modulus is identical. For $\theta_B<\theta<\pi/2$, only the angles $\beta_{1}$ (red) and $\beta_{2}$ (blue) evolve. The  Bloch vectors $\vec{i}_1$ and $\vec{i}_2$ move away from each other on the same longitude, as shown on figure \ref{fig:Protocol}(c-d). In this case, the discriminant square root is a positive real number and the phase of the solutions does not change as both solutions remain positive in this parameter range.

\begin{figure}[t]
	\begin{center}
			\includegraphics[width=1\textwidth]{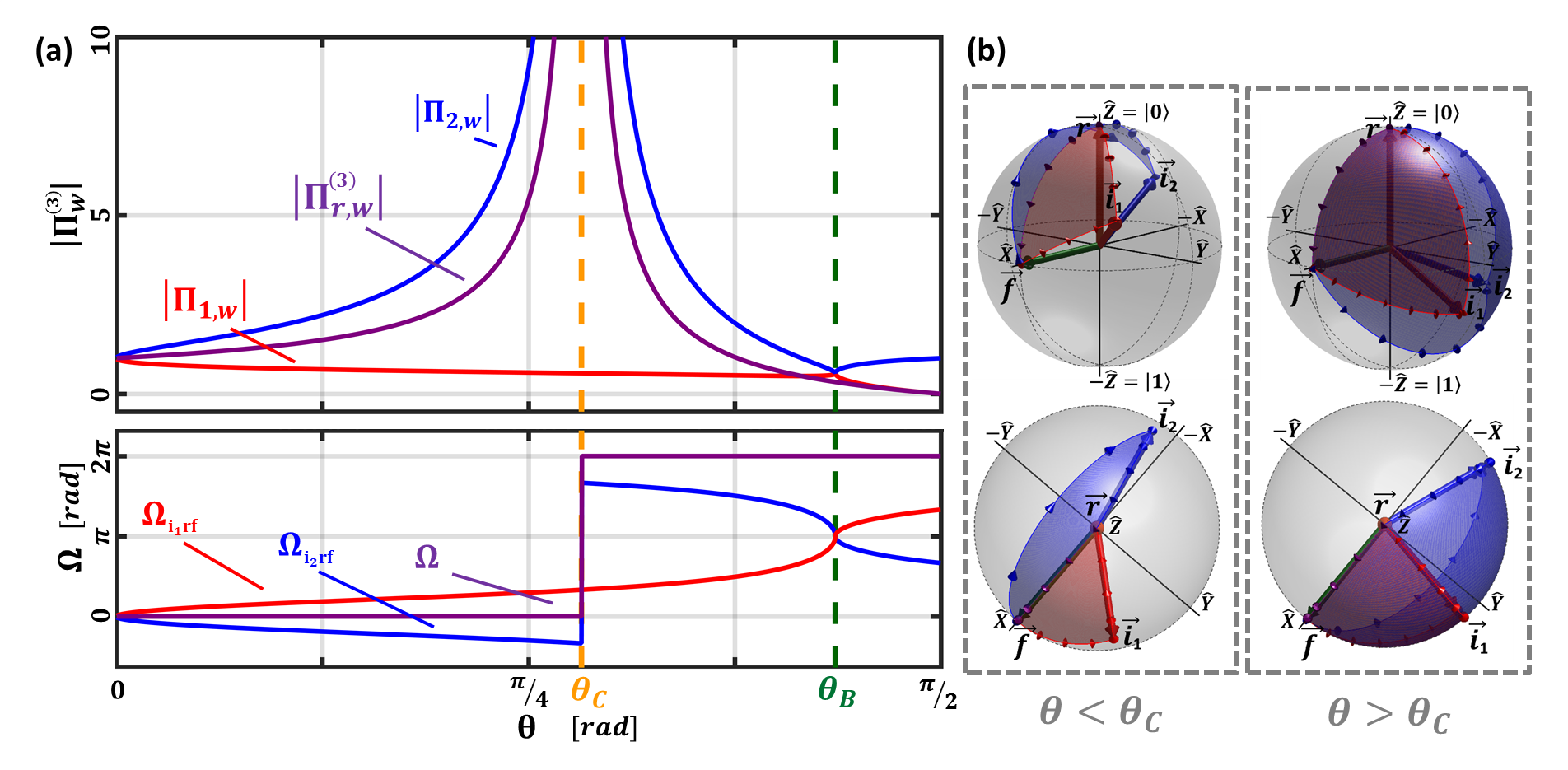} 
            \caption{(a) Modulus of the weak value  $\Pi_{r,w}^{(3)}$  and solid angle defining its argument $\arg\Pi_{r,w}^{(3)}=-\frac{1}{2}\Omega$, as a function of the parameter $\theta$ (violet lines). The weak value of the qutrit projector is related to the weak values of a qubit projector for two different initial states: $|\Pi_{r,w}^{(3)}|=|\Pi_{1,w}|\cdot|\Pi_{2,w}|$ and $\arg\Pi_{r,w}^{(3)}=\arg\Pi_{1,w}+\arg\Pi_{2,w}$. The two initial Bloch vector $\vec{i}_{1}$ and $\vec{i}_{2}$ define independent trajectories on the Bloch sphere as a function of $\theta$ (red and blue lines, respectively). (b) Representation of the solid angles $\Omega_{i_{1}rf}$ and $\Omega_{i_{2}rf}$ (red and blue surface respectively) corresponding to two particular situations, with $\theta<\theta_{C}$ and $\theta>\theta_{C}$.} 
            \label{fig:WeakValueProjector}
	\end{center}
\end{figure}  
\noindent
The weak value $\Pi_{r,w}^{(3)}$ diverges for $\theta_{C}\approx 0.28(2)\pi$ (yellow circles). The argument of the weak value is discontinuous function of $\theta$ as it experiences a $\pi$-phase jump at $\theta_{C}$. The divergence of the weak value occurs when the initial state becomes orthogonal to the final state. In the Majorana representation, at least one of the Bloch vectors describing  the initial state has to be anti-parallel to the postselection vector $\vec{f}=\vec{e}_x$. Figure \ref{fig:Protocol} reveals that the blue, second Bloch vector $\vec{i}_2$ aligns with the $-\vec{e}_x$ direction (azimuth angle $\alpha_2=\pi$ and polar angle $\beta_2=\pi/2$). Hence, the modulus of the weak value $\Pi_{2,w}$ diverges, as seen on figure \ref{fig:WeakValueProjector}(a), and the corresponding solid angle $\Omega_{i_{2}rf}$ is not defined. For the preselected vector $\vec{i}_2$, there exist at least two geodesic trajectories to reach the Bloch vector $\vec{f}$. This induces multiple values for the corresponding solid angle. The origin of the $\pi$-phase jump around this indeterminacy becomes clearer by analyzing the solid angles around the critical angle $\theta_{C}$ on  figure \ref{fig:WeakValueProjector}(b). When $\theta<\theta_{C}$, the  sequence $|\phi_{i}^{(2)}\rangle\rightarrow|\phi_{r}\rangle\rightarrow|\phi_{f}\rangle\rightarrow|\phi_{i}^{(2)}\rangle$ defines a small solid angle in the $XY_{-}X_{-}$ hemisphere. The sequence travels clockwise so that the angle is negative. When $\theta>\theta_{C}$, this sequence runs anti-clockwise and corresponds to a positive solid angle that covers a large part of the $XYX_{-}$ hemisphere. The two situations coexist at $\theta=\theta_{C}$, so that the argument is undefined. This induces a discontinuity of $\Omega_{i_{2}rf}$, which abruptly increases by $2\pi$  across $\theta_C$. The solid angle $\Omega_{i_{1}rf}$ associated with the sequence $|\phi_{i}^{(1)}\rangle\rightarrow|\phi_{r}\rangle\rightarrow|\phi_{f}\rangle\rightarrow|\phi_{i}^{(1)}\rangle$ runs anti-clockwise over the whole parameter range, so that it is always positive. Its value is continuous across the weak value divergence. For $\theta<\theta_C$, the values of the two solid angles are symmetric with respect to 0, while for $\theta>\theta_C$, they are symmetric with respect to $\pi$. Therefore their sum is 0 below $\theta_C$ and $2\pi$ above $\theta_C$.   

\subsection{A well-known quantum paradox in a new form}

In this section, we exploit the Majorana representation to extend the three-box paradox \cite{Aharonov (1991)} to a larger class of quantum phenomena: quantum entanglement. Conceptually, the three-box experiment involves particles that were succesfully pre- and postselected in the three-level quantum states $|\psi_{i}\rangle=\frac{1}{\sqrt{3}}(1,1,1)^{T}$ and $|\psi_{f}\rangle=\frac{1}{\sqrt{3}}(1,-1,1)^{T}$. All other particles are ignored. We define the boxes by the basis states $|\psi_1\rangle=(1,0,0)^{T}$, $|\psi_2\rangle=(0,1,0)^{T}$ and $|\psi_3\rangle=(0,0,1)^{T}$. The three-box paradox deals with the question of determining in which box the particles were between pre- and postselection. To answer this question, as a thought experiment, we can open one or several boxes between pre- and postselection. Then, using the classical rules of conditional probabilities, which in this context are known as the Aharonov-Bergmann-Lebowitz (ABL) rule  \cite{Aharonov (1964)}, we can determine the probability of finding the particle in any box opened. This leads to contradictory conclusions about the intermediate state of the pre- and postselected particles. For example, if we were to open all three boxes simultaneously, we would find the particle in any of the boxes with probability $\frac{1}{3}$. However, if we were to open only box 1 or only box 3, we would find the particle with probability one in the box opened. The ABL-rule is contextual for systems with a Hilbert space $d\geq 3$, i. e. the outcome depends on how the observable was measured. Here the observables are two orthogonal projectors when one box is open but three orthogonal projectors when two or three boxes are opened. The paradoxical nature of the three-box experiment is strongly debated in the literature \cite{Aharonov (1991), Resch (2004), George (2013), Ravon (2007)}. Interestingly, some authors investigated this paradoxical behavior with weak measurements of the box projectors, considering their weak values as non-contextual pseudo-probabilities \cite{Resch (2004)}. It is straightforward to show that the corresponding projector  weak values are $\hat{P}_{1,w}=\hat{P}_{3,w}=1$ and $\hat{P}_{2,w}=-1$.

\begin{figure}[t!]
	\begin{center}
			\includegraphics[width=0.65\textwidth]{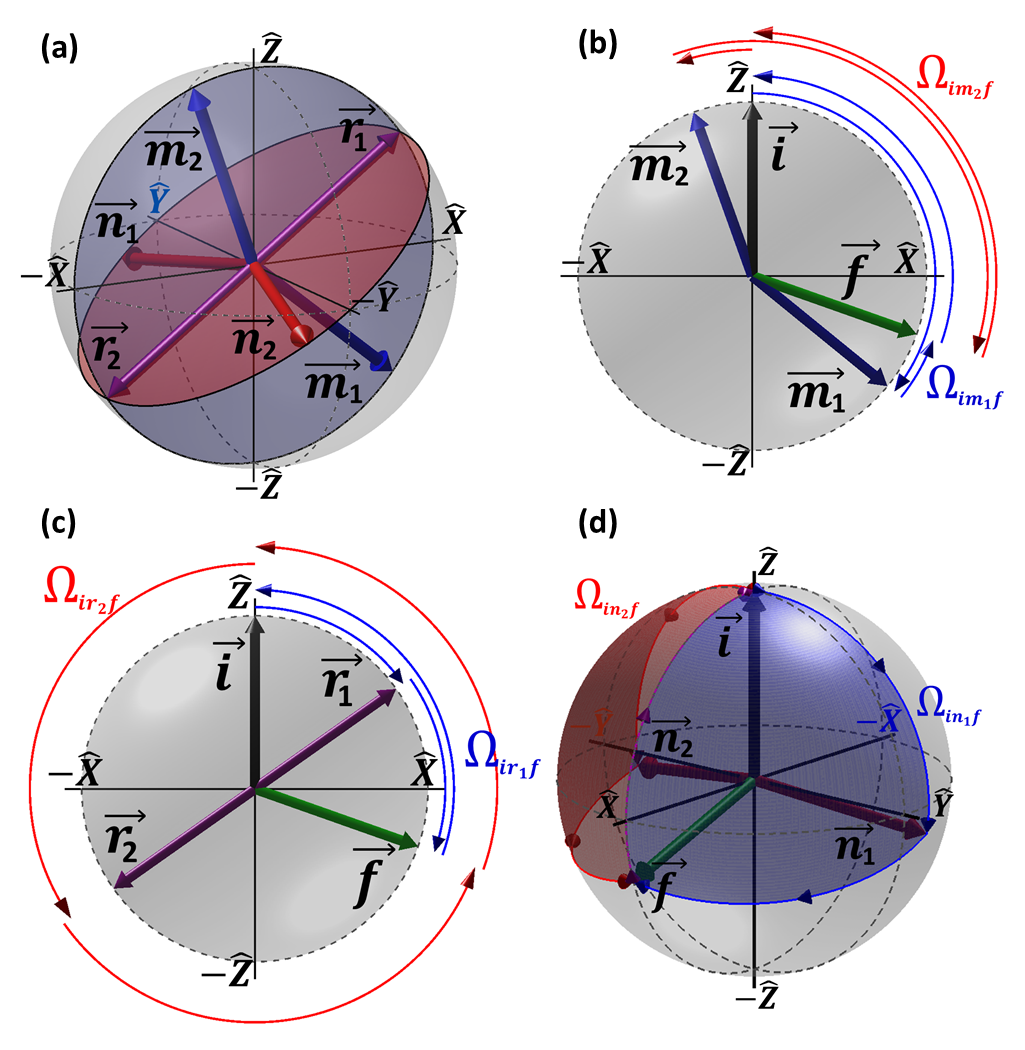} 
            \caption{Representation in the Bloch sphere of the relevant states appearing in the three-box experiment. (a) The three couples of vectors $\vec{m}_{1,2}$, $\vec{r}_{1,2}$ and $\vec{n}_{1,2}$ form a structure which corresponds to the symmetry group $C_{2\nu}$. (b-d) By introducing the pre- and postselected vectors $\vec{i}$ and $\vec{f}$, each solid angle is determined by following the geodesic trajectories.} 
            \label{fig:TBPFigure}
	\end{center}
\end{figure}

We now reformulate this paradox in terms of a bipartite quantum system, using the Majorana representation of all states involved in the experiment. The successive application of the unitary transformations
\begin{equation}
\hat{U}^{(1)}=
\frac{1}{\sqrt{6}}\left( \begin{array}{ccc}
-\sqrt{3} & \sqrt{3} & 0 \\
-1 & -1 & 2 \\
\sqrt{2} & \sqrt{2} & \sqrt{2} \end{array} \right),
\:\:\hat{U}^{(2)}=
\left(\begin{array}{ccc}
\frac{-1-\sqrt{3}}{2\sqrt{2}} & \frac{1-\sqrt{3}}{2\sqrt{2}} & 0 \\
\frac{1-\sqrt{3}}{2\sqrt{2}} & \frac{1+\sqrt{3}}{2\sqrt{2}}& 0 \\
0 & 0 & 1\end{array} \right)\label{eq:TBP Transformation}
\end{equation}
leads to the factorisable  pre- and postselected states $|\Psi_{i}''\rangle=|0\rangle|0\rangle$ and $|\Psi_{f}''\rangle=|\phi_{f}\rangle|\phi_{f}\rangle$ with the appropriate Bloch vectors $\vec{i}=\left(0,0,1\right)$ and $\vec{f}=\frac{1}{3}\left(2\sqrt{2},0,-1\right)$. The resolution of the Majorana polynomial of the three box states transformed under the unitary transformations  (\ref{eq:TBP Transformation}) provides the following three pairs of Bloch vectors:
\begin{eqnarray}
|\Psi_{1}''\rangle \rightarrow \vec{n}_{1,2}=\frac{1}{ \sqrt{3}}\left(-\sqrt{2}\,x,\pm \sqrt[4]{3}\,\sqrt{6\, x},-x\right)\:,\nonumber \\
|\Psi_{2}''\rangle \rightarrow \vec{r}_{1,2}=\pm\frac{1}{ \sqrt{3}}\left(\sqrt{2},0,1\right)\,, \\
|\Psi_{3}''\rangle \rightarrow \vec{m}_{1,2}=\frac{1}{ \sqrt{3}}\left(\pm 2\;\sqrt{x\, (1\pm \sqrt[4]{3}\,\sqrt{ x})},0,x\mp 2\sqrt[4]{3}\, \sqrt{x}\right)\:,\nonumber 
\label{eq:D2dsymmetric}
\end{eqnarray}
where $x=2-\sqrt{3}$. The appropriate normalization factors for the symmetrized states are $K_{r}^{-1}=\sqrt{2}$ and  $K_{m}^{-1}=K_{n}^{-1}=2\sqrt{3}-2$. We represent these six vectors on the Bloch sphere  in figure \ref{fig:TBPFigure}(a), revealing an elegant symmetry. The vectors $\vec{r}_{1}$ and $\vec{r}_{2}$ are anti-parallel and lie at the intersection between the blue and red planes defined by the other pairs of Bloch vectors $\vec{m}_{1,2}$ and $\vec{n}_{1,2}$, respectively. These planes are orthogonal to each other and each plane acts as a mirror plane for the vectors defining the other plane (i. e. the plane containing vectors $\vec{n}_{1}$ and $\vec{n}_{2}$ defines a reflection symmetry between the vectors $\vec{m}_{1}$ and $\vec{m}_{2}$ and the converse symmetry holds as well when exchanging the roles of the pairs). Let us also note that the $\sim74^\circ$ angle between $\vec{r}_1$ and the two vectors $\vec{m}_{1,2}$ is equal to the angle between $\vec{r}_{2}$ and the two vectors $\vec{n}_{1,2}$ (the $\vec{m}_{1,2}$ and $\vec{n}_{1,2}$ pairs are related through a $90^\circ$ rotation-reflection symmetry with respect to the $\vec{r}_{1,2}$ axis). Finally, the vectors $\vec{i}$ and $\vec{f}$ associated with the pre- and postselected states are placed symmetrically around the $\vec{r}_1$ vector in the blue plane defined by the $\vec{m}_{1,2}$ pair, so that they are mirror images of each other with respect to the red symmetry plane. Consequently, the structure formed on the Bloch sphere by all vectors involved in the three-box experiment corresponds to the symmetry group $C_{2\nu}$. 

This symmetry allows us to introduce the local rotation operation $\hat{U}_r=-\hat{\sigma}_{r}\otimes\hat{\sigma}_{r}$ under which the three-box experiment is left invariant (the operator $\hat{\sigma}_r=\vec{r}_1\cdot\vec{\hat{\sigma}}$ effectively exchanges the vectors $\vec{m}_{1}$ with $\vec{m}_{2}$,  $\vec{n}_{1}$ with $\vec{n}_{2}$, and  $\vec{i}$ with $\vec{f}$, while leaving  $\vec{r}_{1,2}$ invariant; note that states may pick up a phase in the process). In particular, $\hat{\sigma}_{r}|0\rangle=|\phi_{f }\rangle$. Therefore, the weak values of the box projectors are necessarily real, while their argument of either $0$ or $\pi$ determines their sign. Indeed, the weak value on the $k^{th}$ box consist of the products of two projector weak values $\Pi_{k,w}^{(3)}=\Pi_{k_1,w}^{(2)}\Pi_{k_2,w}^{(2)}$. For the projectors on box two and three, all vectors are in the same (blue) plane. As result, the solid angles determining the argument of the weak values can take only the values 0 or $2 \pi$, as shown figure \ref{fig:TBPFigure}(b-c), so that they determine the sign of the weak values. To show that the first projector takes are real value, we apply the unitary transformation $\hat{\sigma}_r$ associated with the symmetry: $\Pi_{n_1,w}^{(2)}=\langle\phi_{f}|\hat{\sigma}_r^\dagger\hat{\sigma}_r| \phi_{n_1}\rangle \langle\phi_{n_1}|\hat{\sigma}_r^\dagger\hat{\sigma}_r| \phi_{i}\rangle \langle\phi_{f}|\hat{\sigma}_r^\dagger\hat{\sigma}_r| \phi_{i}\rangle^{-1}=\langle\phi_{i}| \phi_{n_2}\rangle \langle\phi_{n_2}| \phi_{f}\rangle \langle\phi_{i}| \phi_{f}\rangle^{-1}=\Pi_{n_2,w}^{(2)*}$, so that the two weak values are complex conjugates of each other. The corresponding solid angles are shown on  figure \ref{fig:TBPFigure}(d). By applying the general relations (\ref{eq:GeneralModulus}) and (\ref{eq:GeneralSolidAngle}), it is straightforward to show, that the values determined by the geometric approach are in agreement with the standard results of the quantum three-box paradox (see Table \ref{Table:WeakValues3box}). Here, we see that the negative sign of the weak value of the second box projector $\hat{P}_{2,w}$ arises from the quantum geometric phase of $\pi$, which emphasizes its non-classical origin. 

\begin{table}[h]
\centering
\renewcommand{\arraystretch}{1.5}
\begin{tabular}{lll}
\hline
\textbf{Box 1}   \,\,\,\,\,\, & $|\Pi_{n_{1},w}|=1$ & \,\,\,\,\,\,$\Omega_{in_{1}f}=-2\,\arctan\sqrt{3+2\, \sqrt{3}}$\\
$\hat{P}_{1,w}=1  \,\,\,\,\,\, $& $|\Pi_{n_{2},w}|=1$ & \,\,\,\,\,\,$\Omega_{in_{2}f}=+2\,\arctan\sqrt{3+2\,\sqrt{3}}$\\
\hline
\textbf{Box 2}   \,\,\,\,\,\, & $|\Pi_{r_{1},w}|=\sqrt{2+\sqrt{3}}$ & \,\,\,\,\,\,$\Omega_{ir_{1}f}=0$\\
$\hat{P}_{2,w}=-1  \,\,\,\,\,\, $ & $|\Pi_{r_{2},w}|=\sqrt{2-\sqrt{3}}$ & \,\,\,\,\,\,$\Omega_{ir_{2}f}=2\pi$ \\ 
\hline
\textbf{Box 3}  \,\,\,\,\,\, & $|\Pi_{m_{1},w}|=1$ & \,\,\,\,\,\,$\Omega_{im_{1}f}=0$ \\ 
$\hat{P}_{3,w}=1  \,\,\,\,\,\, $ & $|\Pi_{m_{2},w}|=1$ & \,\,\,\,\,\,$\Omega_{im_{2}f}=0$\\ 
\hline
\end{tabular}  
\caption{Weak values of the box projectors in the three-box paradox determined from the weak values of the associated qubit projectors deduced from the Majorana representation.}\label{Table:WeakValues3box}
\end{table} 

We now consider the physical interpretation of the box projectors in the two-qubit space. Due to the rotational invariance of the problem, the orthogonal box states are orthogonal eigenstates of the rotation operator $\hat{U}_{r}$. Because the qubit states associated with the vectors $\vec{r}_{1,2}$ are orthogonal, they define a basis of the qubit Hilbert space, noted $\left\{|\phi_{r}\rangle,| \phi_{-r}\right\}$ where the actual states are $|\phi_{\pm r}\rangle=\sqrt{\frac{1}{6}(3 \pm \sqrt{3})}\: |0\rangle \pm \sqrt{\frac{1}{6}(3 \mp \sqrt{3})}\: |1\rangle$. Using this basis to express the box states in the Majorana representation, we find that the relevant four orthogonal eigenstates of the operator are:
\begin{eqnarray}
|\Psi_{1}''\rangle&=&\frac{|\phi_{r}\rangle|\phi_{r}\rangle+\sqrt{3} |\phi_{-r}\rangle|\phi_{-r}\rangle}{2},  \label{eq:3box-box1} \\
|\Psi_{2}''\rangle&=&\frac{|\phi_{r}\rangle|\phi_{-r}\rangle+|\phi_{-r}\rangle|\phi_{r}\rangle}{\sqrt{2}},      \label{eq:3box-box2} \\
|\Psi_{3}''\rangle&=&\frac{\sqrt{3}|\phi_{r}\rangle|\phi_{r}\rangle-|\phi_{-r}\rangle|\phi_{-r}\rangle}{2},   \label{eq:3box-box3}\\
|\widetilde{\Psi}_{4}''\rangle&=&\frac{|\phi_{r}\rangle|\phi_{-r}\rangle-|\phi_{-r}\rangle|\phi_{r}\rangle}{\sqrt{2}}, \label{eq:3box-box4}
\end{eqnarray}
where $|\Psi_{2}''\rangle$ and $|\widetilde{\Psi}_{4}''\rangle$ are associated with the eigenvalue $+1$ and where $|\Psi_{1}''\rangle$ and $|\Psi_{3}''\rangle$ are associated with the eigenvalue $-1$. The state $|\Psi_{2}''\rangle$ above defines box 2 because it matches obviously the symmetrized form obtained from the Majorana representation in terms of the states associated with the vectors $\vec{r}_{1,2}$. It corresponds to a maximally entangled Bell state.  The eigenvector $|\widetilde{\Psi}_{4}''\rangle$  shares its eigenvalue with $|\Psi_{2}''\rangle$ but cannot represent a state of the three-level system because it corresponds to the anti-symmetric subspace of the two-qubit space. Therefore, the states representing the box-1 and box-3 projectors are necessarily orthogonal vectors in the subspace spanned by the two-qubit states $|\phi_{r}\rangle|\phi_{r}\rangle$ and $|\phi_{-r}\rangle|\phi_{-r}\rangle$, which share the same eigenvalue $-1$ of $\hat{U}_r$. The calculations leading to the exact form (\ref{eq:3box-box1}) and (\ref{eq:3box-box3}) of states $|\Psi_{1}''\rangle$ and  $|\Psi_{3}''\rangle$ are explained in \ref{Orthogonal-states decomposition}. The states are non-maximally entangled, with a von Neumann entropy of $0.81$ (the von Neumann entropy is $0$ for pure states and $1$ for maximally entangled 2-qubit states). The degree of entanglement of these states can also be ascertain geometrically on the Bloch sphere \cite{Usha (2012), Aulbach (2010), Aulbach (2011)}, by looking at the angle between the two vectors representing the symmetric state (antipodal Majorana points correspond to maximally entangled states while superposed Majorana points correspond to separable states). Let us note as well that the closest separable state is given by the angle bisector between the two Majorana points \cite{Aulbach (2011)}. These corresponds to the state $ |\phi_{-r}\rangle|\phi_{-r}\rangle$ for box-1 state $|\Psi_{1}''\rangle$ and to the orthogonal state  $|\phi_{r}\rangle|\phi_{r}\rangle$ for box-3 state $|\Psi_{3}''\rangle$.

Now we can reformulate the three-box paradox in terms of the two-particle system. We consider all --and only-- the particles that were successfully pre- and postselected in the separable states $|\Psi_{i}''\rangle=|0\rangle|0\rangle$ and $|\Psi_{f}''\rangle=|\phi_{f}\rangle|\phi_{f}\rangle$, respectively. We define the boxes by projective measurements on the states $|\Psi_{1}''\rangle$, $|\Psi_{2}''\rangle$, $|\Psi_{3}''\rangle$, and $|\widetilde{\Psi}_{4}''\rangle$. Note that we can safely ignore the fourth box defined by the projector on the antisymmetric state $|\widetilde{\Psi}_{4}''\rangle$ because it is orthogonal to both the initial and final symmetric states. This experiment corresponds to a Bell-type measurement, related to the unitary observable $\hat{U}_{r}=-\hat{\sigma}_{r}\otimes\hat{\sigma}_{r}$. This observable is generally used to point out non-classical correlations between bipartite qubit systems. Exactly as in the standard formulation of the paradox, if we were to open box one or box three, we would find the particle there with certainty. In this case however, we would deduce that the particles are necessarily entangled, although both their initial and final states are classical separable states. In the standard formulation of the paradox, the particles start and end in a superposed quantum state and the measurement is represented by three classical boxes. After transposing the paradox in the two-qubit Majorana representation, our particles start and end in a classical state but our boxes become quantum and entangle the particles in the process. The occurrence of entanglement in the bipartite system is an unavoidable feature of the Majorana representation of the paradox. In particular, it cannot be removed through a unitary transformation because one of the basis state of the three-level system is necessarily entangled in the symmetric two-qubit representation and any attempt to disentangle the qubit states associated with the box projectors would entangle the qubit states describing the initial and final states.

\section{Conclusion}

In this paper, we expressed the polar form of weak and modular values of operators of two-, three- and higher-level systems that describe preselected and postselected experiments, such as weak measurements involving discrete quantum systems. We used the Majorana representation of $N$-level systems, which assigns a correspondance between states of $N$-level systems and symmetric states of $N-1$ qubits. This new approach led to a geometric description of weak and modular values in terms of vectors on the Bloch sphere. We found that weak values of projectors and modular values can be factored in $N-1$ contributions by considering the underlying qubit contributions associated with the Majorana representation. Their modulus is determined by a product of $N-1$ square roots involving ratios of projection probabilities between qubit states. The latter are expressed as a function of scalar products between Bloch vectors. Their argument is given by a sum of $N-1$ half solid angles related to $N-1$ spherical polygons defined by qubit states on the Bloch sphere. The arguments of weak and modular values correspond thus to a quantum geometric phase. Their values are expressed as a function of scalar and cross products between Bloch vectors. Exploiting this geometric approach, we studied the discontinuities of the argument of the weak value of a three-level projector, which occur around singularities of the weak value for orthogonal pre- and postselected states. We found its origin in the geometric phase, as one of its contributing solid-angle jumps by $2\pi$ across the singularity, which translates in a $\pi$-phase jump in the weak value (equivalent to a sign flip). Moreover, this theoretical framework based on the Majorana representation allowed us to recast the three-box paradox in a new form, which involves quantum entanglement. We analyzed the weak values of the box projectors in terms of vectors on the Bloch sphere. We found that the origin of the negative sign occurring in  one the weak values -- which has been sometimes interpreted has a $-1$ pseudo-probability in the literature -- is directly related to a geometric quantum phase defined on the Bloch sphere. In the two-particle version of the three-box paradox, the particles are pre- and postselected in classical separable states but are necessarily found in entangled intermediate states when opening one amongst two of the three boxes. In this representation of the paradox, the boxes are quantum, represented by projectors on eigenvectors of a Bell-type measurement observable, while the initial and final states are classical. A paradoxical formulation of this observation would pose the question of the classical vs quantum evolution of the particle pairs in the pre- and postselected ensemble. The two applications that we explored within the framework of the geometric description on the Bloch sphere show the usefulness of our method using the Majorana representation of the $N$-level system. As this description is general, it should prove useful for all problems involving weak and modular values of discrete quantum systems.

\section*{Acknowledgments}
Y. C. is a research associate of the Belgian Fund for Scientific Research F.R.S.-FNRS. The authors would like to thank  M. Lobet,  K. Wilden, A. Reul and B. Kolaric for their careful reading of the paper and their insightful remarks.

\appendix

\section{Weak and modular values of qubit observables expressed using Bloch vectors}
\label{Modulus Calculation}

\subsection{Modulus expression}
\label{app:modulus expression}

The projection probability between two arbitrary qubit states  $\left|\langle\phi_{v}|\phi_{u}\rangle\right|^2$ is given by the trace $\Tr \hat{\Pi}_{v}\hat{\rho}_u$ where the projector is $\hat{\Pi}_{v}=\frac{1}{2}(\hat{I}+\vec{v}\cdot\vec{\hat{\sigma}})$ and the density operator is equivalently expressed by $\hat{\rho}_u=\frac{1}{2}(\hat{I}+\vec{u}\cdot\vec{\hat{\sigma}})$. Products between Pauli matrices verify the well-known property \cite{Kofman (2012)}
\begin{equation}\label{eq:pauliscalarprod}
(\vec{v}\cdot\vec{\hat{\sigma}})(\vec{u}\cdot\vec{\hat{\sigma}})=(\vec{v}\cdot\vec{u}) \:\hat{I}+j\:(\vec{v}\times\vec{u})\cdot \vec{\hat{\sigma}}
\end{equation}
due to their commutation rules, where $j$ is the unit imaginary number. The operator to be traced is thus given by
\begin{equation}\label{eq:projprobabeforetrace}
\hat{\Pi}_{v}\hat{\rho}_{u}=\frac{1}{4}\left(1+\vec{v}\cdot\vec{u}\right)\hat{I}+\frac{1}{4}\left[\,\vec{u}+\vec{v}+j\,(\vec{v}\times\vec{u})\,\right]\cdot\vec{\hat{\sigma}}\,.
\end{equation}
When taking the trace, only the first term survives because Pauli matrices are traceless. Thus, the projection probability is equal to
\begin{equation}\label{eq:projproba}
\left|\langle\phi_{v}|\phi_{u}\rangle\right|^2=\Tr\left[\hat{\Pi}_{v}\hat{\rho}_{u}\right]=\frac{1}{2}\left(1+\vec{v}\cdot\vec{u}\right)\,.
\end{equation}
Since weak and modular values are given by products and ratios of state overlaps through expressions (\ref{projectorweakvaluedef}) and (\ref{2lvlmodularvaluedef}), their modulus take the form of products and ratios of square roots of the form $\left|\langle\phi_{v}|\phi_{u}\rangle\right|=\sqrt{\frac{1}{2} (1+\vec{v}\cdot\vec{u})}$.

\subsection{Qubit projection Operator}

The weak value of a qubit projector is by definition \cite{Kofman (2012)}
\begin{equation}
\hat{\Pi}_{r,w}=\frac{\langle \phi_f|\hat{ \Pi}_r|\phi_i\rangle}{\langle \phi_f | \phi_i \rangle}=\frac{\Tr\left[\hat{\Pi}_{f}\,\hat{\Pi}_{r}\hat{\rho}_{i}\right]}{\Tr\left[\hat{\Pi}_{f}\hat{\rho}_{i}\right]}\,.
\label{eq:WeakValueDefinition}
\end{equation}
The denominator is given by expression (\ref{eq:projproba}) with the appropriate substitutions $\vec{u}=\vec{i}$ and $\vec{v}=\vec{f}$. To find the numerator, we start from result (\ref{eq:projprobabeforetrace}) but with the substitutions $\vec{u}=\vec{i}$ and $\vec{v}=\vec{r}$, so that
\begin{eqnarray}
\hat{\Pi}_{f}\hat{\Pi}_{r}\hat{\rho}_{i}& = &\frac{1}{4}\left(1+\vec{r}\cdot\vec{i}\right)\hat{\Pi}_f+\frac{1}{4}\hat{\Pi}_f\left[\,\vec{i}+\vec{r}+j\,(\,\vec{r}\times\vec{i}\;)\,\right]\cdot\vec{\hat{\sigma}}\\
  & = & \frac{1}{8}\left(1+\vec{r}\cdot\vec{i}\right)\left(\hat{I}+\vec{f}\cdot\vec{\hat{\sigma}}\right)+\frac{1}{8}\left(\hat{I}+\vec{f}\cdot\vec{\hat{\sigma}}\right)\left[\vec{i}+\vec{r}+j\, (\,\vec{r}\times\vec{i}\;)\right]\cdot\vec{\hat{\sigma}}\nonumber \, ,
\end{eqnarray}
where we replaced the projector $\hat{\Pi}_{f}=\frac{1}{2}(\hat{I}+\vec{f}\cdot\vec{\hat{\sigma}})$ by its expression. Using property (\ref{eq:pauliscalarprod}) to resolve the product between the Pauli matrices, this expression expands to
\begin{eqnarray}\label{eq:numeratortobetraced}
\hat{\Pi}_{f}\hat{\Pi}_{r}\hat{\rho}_{i}
&=& \frac{1}{8}\left(1+\vec{r}\cdot\vec{i}\right)\left(\hat{I}+\vec{f}\cdot\vec{\hat{\sigma}}\right) + \frac{1}{8}\left[\vec{i}+\vec{r}+j\, (\,\vec{r}\times\vec{i}\;)\right]\cdot\vec{\hat{\sigma}} \\
 & + &  \frac{1}{8}\; \vec{f} \cdot \left[\vec{i}+\vec{r}+j\, (\,\vec{r}\times\vec{i}\;)\right] \hat{I}
     + \frac{1}{8} \,j\left\{ \vec{f}\times\left[\vec{i}+\vec{r}+j\, (\,\vec{r}\times\vec{i}\;)\right]\right\}\cdot\vec{\hat{\sigma}}  \nonumber . 
\end{eqnarray}
Taking the trace of this expression suppresses all the terms involving Pauli matrices, so that the weak value of the projector $\hat{\Pi}_{r}$  is finally given by:
\begin{equation}\label{eq:qubitprojectweakvalueonBloch}
\Pi_{r,w}=\frac{\Tr\left[\hat{\Pi}_{f}\,\hat{\Pi}_{r}\hat{\rho}_{i}\right]}{\Tr\left[\hat{\Pi}_{f}\hat{\rho}_{i}\right]}=\frac{1}{2}\frac{1+\vec{f}\cdot\vec{r}+\vec{r}\cdot\vec{i}+\vec{f}\cdot\vec{i}+j\,\left[\vec{f}\cdot(\,\vec{r}\times\vec{i}\:)\right]}{1+\vec{f}\cdot\vec{i}}\:.
\end{equation}
The argument of the weak value given by (\ref{eq:argprojwvalqubit}) is deduced immediately from this expression by considering the real and the imaginary part of the numerator (proper care should be given to the sign of the numerator and denominator in the arctangent function to determine the correct quadrant of the angle). 

\subsection{Qubit unitary operator}\label{Qubit unitary operator}

The modular value $\sigma^{\alpha,\beta}_{r,m}$ of the qubit unitary operator $\hat{U}_{\sigma_{r}}^{\alpha,\beta}=e^{j\frac{\beta}{2}}e^{-j\frac{\alpha}{2}\hat{\sigma}_{r}}$ is defined by 
\begin{equation}
\sigma^{\alpha,\beta}_{r,m}=e^{j\frac{\beta}{2}}\,\frac{\langle \phi_f|e^{-j\frac{\alpha}{2}\hat{\sigma}_{r}}|\phi_i\rangle}{\langle \phi_f | \phi_i \rangle}=e^{j\frac{\beta}{2}}\,\frac{\Tr\left[\hat{\Pi}_{f}\,e^{-j\frac{\alpha}{2}\hat{\sigma}_{r}}\hat{\rho}_{i}\right]}{\Tr\left[\hat{\Pi}_{f}\hat{\rho}_{i}\right]}\,.
\label{eq:ModularValueDefinition}
\end{equation}
The denominator is given by expression (\ref{eq:projproba}) with the appropriate substitutions $\vec{u}=\vec{i}$ and $\vec{v}=\vec{f}$. Considering that the Pauli operator can be expressed as the difference between two orthogonal projectors $\hat{\sigma}_{r}=\hat{\Pi}_{r}-\hat{\Pi}_{-r}$, we can write the numerator as: 
\begin{equation}\label{eq:modvalnumqubit}
 \Tr\left[\hat{\Pi}_{f}e^{-j\frac{\alpha}{2}\hat{\sigma}_{r}}\hat{\rho}_{i}\right]= e^{-j\frac{\alpha}{2}}\,\Tr\left[\hat{\Pi}_{f}\hat{\Pi}_{r}\hat{\rho}_{i}\right]+e^{j\frac{\alpha}{2}}\,\Tr\left[\hat{\Pi}_{f}\hat{\Pi}_{-r}\hat{\rho}_{i}\right] .
\end{equation}
The calculation of the first trace was already performed in expressions (\ref{eq:numeratortobetraced}--\ref{eq:qubitprojectweakvalueonBloch}), while the second trace can be obtained from this previous result by replacing the vector $\vec{r}$ by $-\vec{r}$. Therefore, we find that the numerator (\ref{eq:modvalnumqubit}) becomes
\begin{eqnarray}
 \frac{e^{-j\frac{\alpha}{2}}}{4}\left(1+\vec{f}\cdot\vec{r}+\vec{r}\cdot\vec{i}+\vec{f}\cdot\vec{i}+j\,V\right)\nonumber 
+\frac{e^{j\frac{\alpha}{2}}}{4}\left(1-\vec{f}\cdot\vec{r}-\vec{r}\cdot\vec{i}+\vec{f}\cdot\vec{i}-j\,V\right) \nonumber \\
= \frac{1}{2}\left\{\cos\frac{\alpha}{2}\left(1+\vec{f}\cdot\vec{i}\right)+ \sin\frac{\alpha}{2}\left[V-j\, (\,\vec{f}\cdot\vec{r}+\vec{r}\cdot\vec{i}\;)\right]\right\} .
\end{eqnarray}
where we wrote the signed volume of the parallelipiped defined by the vector triad by $V=\vec{f}\cdot (\,\vec{r}\times\vec{i}\;)$. In the end, we obtain the following expression for the modular value as a function of Bloch vectors:  
\begin{equation}
\sigma^{\alpha,\beta}_{r,m}=e^{j\frac{\beta}{2}}\frac{\cos\frac{\alpha}{2}\left(1+\vec{f}\cdot\vec{i}\right)+ \sin\frac{\alpha}{2}\left[V-j\,(\,\vec{f}\cdot\vec{r}+\vec{r}\cdot\vec{i}\;)\right]}{1+\vec{f}\cdot\vec{i}}\,.\nonumber
\label{eq:ModularValueQubitREL}
\end{equation}
The total argument of the modular value can be readily deduced from the expression above. It contains a dynamical contribution $(\beta - \alpha)/2$ and a geometrical contribution defined by $\Omega=-\Omega_{irsf}/2$. We now evaluate the geometrical contribution $\Omega$ in terms of Bloch vectors:
\begin{equation}
\Omega = \arg \left\{\left\{\cos\frac{\alpha}{2}\left(1+\vec{f}\cdot\vec{i}\right)+ \sin\frac{\alpha}{2}\left[V-j\,(\,\vec{f}\cdot\vec{r}+\vec{r}\cdot\vec{i}\;)\right]\right\}e^{j\frac{\alpha}{2}}\right\}\, ,
\end{equation}
where the phase factor at the end is required to remove the appropriate dynamical contribution. By expanding this expression, we find the value of the geometric phase as a function of Bloch vectors:
\begin{eqnarray}\label{eq:gmphasemodularirsf}
\Omega  = &\arg \biggr\{\enskip&\left[1 + \vec{f}\cdot\vec{i} + V \tan\frac{\alpha}{2} + (\,\vec{f}\cdot\vec{r}+\vec{r}\cdot\vec{i}\;) \tan^2\frac{\alpha}{2} \right]  \\
                  &\enskip\enskip+ j &\left[1+ \vec{f}\cdot\vec{i} + V \tan\frac{\alpha}{2} - (\,\vec{f}\cdot\vec{r}+\vec{r}\cdot\vec{i}\;)  \right] \tan\frac{\alpha}{2}\enskip \biggr\} \;.\nonumber
\end{eqnarray}
This solid angle can be expressed as the sum of two contributions $\Omega=\Omega_1+\Omega_2$, where
\begin{eqnarray}
\Omega_1=&\arg \biggr\{\enskip
\left[ 1+ \vec{f}\cdot\vec{i} +V \tan\frac{\alpha}{2}+(\,\vec{f}\cdot\vec{r}+\vec{r}\cdot\vec{i}\;) \tan^2\frac{\alpha}{2}\;  (\,\vec{r}\cdot\vec{i}\;)  \right] \\
&\enskip\enskip+ j \left[  (\,\vec{f}\cdot\vec{i}\;)\,(\,\vec{r}\cdot\vec{i}\;) +V \tan\frac{\alpha}{2}\; (\,\vec{r}\cdot\vec{i}\;)   - \vec{f}\cdot\vec{r}\;  \right] \tan\frac{\alpha}{2} \enskip\biggr\} \; , \nonumber
\end{eqnarray}
\begin{equation}
\Omega_2=\arg \left\{ \, \left[1+\vec{r}\cdot\vec{i}\;\tan^2\frac{\alpha}{2} \right]    +j \left[ \tan\frac{\alpha}{2}\;(1-\vec{r}\cdot\vec{i}\;)  \right]      \right\}.
\end{equation}
Using a symbolic computation package, it is straightforward to show that $\tan \Omega = \tan (\Omega_1 + \Omega_2)=(\tan\Omega_1+\tan \Omega_2 )/(1-\tan \Omega_1 \tan \Omega_2)$ and that the angles are defined in the proper quadrants. The values given above for $\Omega_1$ and $\Omega_2$ result directly from the definitions of 
\begin{eqnarray}
\Omega_1= &\arg \left[1+\vec{f}\cdot\vec{s}+\vec{s}\cdot\vec{i} + \vec{f}\cdot\vec{i}+j\;\vec{f}\cdot (\,\vec{s}\times\vec{i}\;) \right] =-\frac{1}{2} \Omega_{isf} \;, \\
\Omega_2= &\arg \left[1+\vec{s}\cdot\vec{r}+\vec{r}\cdot\vec{i} + \vec{s}\cdot\vec{i}+j\;\vec{s}\cdot  (\,\vec{r}\times\vec{i}\;) \right]=-\frac{1}{2} \Omega_{irs}\;  ,
\end{eqnarray}
where the vector $\vec{s}$ was expressed by Rodrigue's rotation formula (\ref{eq:RodrigueRotFormula}). As a result, the geometrical phase is related to the solid angle by $\Omega=-\frac{1}{2} (\Omega_{isf}+\Omega_{irf})=-\frac{1}{2} \Omega_{irsf}$. An alternative method to demonstrate this result can be found in \cite{Cormann (2015)} (for the particular case $\alpha =\pi$).

\section{Argument of the modular value for $N$-level quantum systems}
\label{App:argModValqutrit}
We define the traceless Hermitian operator $\hat{\Lambda}_r$ acting on the $N$-level system and express it in the basis of its eigenvectors $\hat{\Lambda}_r=\sum_{r'=1}^N \Lambda_{r'} \vert \psi_{r'}\rangle\langle \psi_{r'}\vert $. The associated unitary operator is
\begin{equation}\label{eq:Udef}
 \hat{U}_{\Lambda_{r}}^{\alpha,\beta}=e^{j \beta}e^{-j \alpha \frac{N-1}{2} \hat{\Lambda}_{r}}=e^{j \beta} \sum_{r'=1}^N e^{-j \alpha \frac{N-1}{2}  \Lambda_{r'}} \vert \psi_{r'}\rangle\langle \psi_{r'}\vert\,.
 \end{equation}
 We define the initial state $\vert\psi_i\rangle$ and write it in the basis of the eigenvectors of $\hat{\Lambda}_r$, so that $\vert\psi_i\rangle= \sum_{r'=1}^N \langle \psi_{r'}\vert \psi_{i}\rangle \,\vert \psi_{r'}\rangle$. We also define the state $\vert \psi_S\rangle$ that result from applying the unitary operator to the initial state:
 \begin{equation}\label{eq:psisdef}
 \vert \psi_S\rangle = \hat{U}_{\Lambda_{r}}^{\alpha,\beta}\vert\psi_i\rangle=e^{j \varphi_s} \vert\psi_s\rangle\, ,
 \end{equation}
 where $\vert\psi_s\rangle $ corresponds to the state $\vert \psi_S\rangle$ written in its cannonical form, i. e. without the global phase factor $\varphi_s$ that it may have acquired under the unitary transformation. To evaluate the phase $\varphi_s$, we project the state $\vert\psi_S\rangle$ on an arbitrary eigenvector $\vert\psi_r\rangle $ of the unitary operator:
\begin{equation}\label{eq:projpsiS}
\langle \psi_r\vert\psi_S\rangle=e^{j \varphi_s} \langle \psi_r\vert\psi_s\rangle=e^{j \beta} e^{-j \alpha \frac{N-1}{2}  \Lambda_{r}} \langle \psi_{r}\vert\psi_i\rangle\,,
\end{equation}
where the first equality results from (\ref{eq:psisdef}) and the second from (\ref{eq:Udef}) and the definition of $\vert\psi_S\rangle$. Equation (\ref{eq:projpsiS}) shows that the projections $\vert \langle \psi_r\vert\psi_s\rangle\vert$ and $\vert \langle \psi_r\vert\psi_i\rangle\vert$ are identical, which is due to the unitary character of the operator $ \hat{U}_{\Lambda_{r}}^{\alpha,\beta}$. By equating the arguments of both sides of equality (\ref{eq:projpsiS}), we find the value of the phase $\varphi_s$:
\begin{equation}
\varphi_s=\beta-\alpha\frac{N-1}{2}\Lambda_r+\arg\langle \psi_r\vert\psi_i\rangle -\arg\langle \psi_r\vert\psi_s\rangle\,.
\end{equation}
The modular value is given by $\Lambda_{r,m}^{\alpha,\beta}=\langle \psi_f\vert\psi_S\rangle \langle \psi_f\vert\psi_i\rangle ^{-1}$. Therefore its argument is given by
\begin{equation}
\arg \Lambda_{r,m}^{\alpha,\beta}=\varphi_s +\arg\langle \psi_f\vert\psi_s\rangle -\arg\langle \psi_f\vert\psi_i\rangle\,.
\end{equation}
Now, we apply the unitary transformation that maps the intial state $\vert \psi_i \rangle $ and the eigenvector state $\vert \psi_r \rangle $ to factored states in the Majorana representation:
\begin{eqnarray}
|\Psi_{i}''\rangle=K_i \sum_{P}\hat{P}\left[|\phi_{i_1}\rangle|\phi_{i_2}\rangle...|\phi_{i_{N-1}}\rangle\right]\: ,\nonumber  \\
|\Psi_{s}''\rangle=K_s \sum_{P}\hat{P}\left[|\phi_{s_1}\rangle|\phi_{s_2}\rangle...|\phi_{s_{N-1}}\rangle\right]\: , \\
|\Psi_{r}''\rangle=\underbrace{|\phi_{r}\rangle...|\phi_{r}\rangle}_{N-1}\:,\:\:\:\:\:\:\:\: 
|\Psi_{f}''\rangle=\underbrace{|\phi_{f}\rangle...|\phi_{f}\rangle}_{N-1}\:. \nonumber
\end{eqnarray}
This transformation leaves invariant the argument of the modular value, so that, ignoring for now the global phase $\beta-\alpha\frac{N-1}{2}\Lambda_r$, the geometrical component becomes
\begin{equation}
\varphi_g=\arg\langle \Psi_f''\vert\Psi_s''\rangle -\arg\langle \Psi_f''\vert\Psi_i''\rangle +\arg\langle \Psi_r''\vert\psi_i''\rangle -\arg\langle \Psi_r''\vert\Psi_s''\rangle\,.
\end{equation}
We can rewrite this phase into two components $\varphi_g=\varphi_{g_1}+\varphi_{g_2}$ defined by
\begin{eqnarray}
\varphi_{g_1}=\arg\langle \Psi_f''\vert\Psi_s''\rangle+\arg\langle \Psi_i''\vert\Psi_f''\rangle\,, \\
\varphi_{g_2}=\arg\langle \Psi_r''\vert\Psi_i''\rangle+\arg\langle \Psi_s''\vert\Psi_r''\rangle\, ,
\end{eqnarray}
where we used the property $\arg\langle \Psi_a\vert\Psi_b\rangle=-\arg\langle \Psi_b\vert\Psi_a\rangle$. In terms of the qubits defining the Majorana representation, these phase components become
\begin{eqnarray}
\varphi_{g_1}'=\sum_{k=1}^{N-1}\left(\arg\langle \phi_f\vert\phi_{s_k}\rangle+\arg\langle \phi_{s_k}\vert\phi_{i_k}\rangle+\arg\langle \phi_{i_k}\vert\phi_{f}\rangle\right)\,, \label{eq:phi1'}\\
\varphi_{g_2}'=\sum_{k=1}^{N-1}\left(\arg\langle \phi_r\vert\phi_{i_k}\rangle+\arg\langle \phi_{i_k}\vert\phi_{s_k}\rangle+\arg\langle \phi_{s_k}\vert\phi_r\rangle\right)\, , \label{eq:phi2'}
\end{eqnarray}
where the middle terms that were added to both equations compensate each other so that $\varphi_g=\varphi_{g_1}'+\varphi_{g_2}'$. Each triplet of arguments summed in (\ref{eq:phi1'}) corresponds to the argument of the weak value $\langle \phi_f\vert \hat{\Pi}_{s_k}\vert\phi_{i_k}\rangle\langle \phi_f\vert \phi_{i_k}\rangle^{-1}$ of a qubit projector on the state $\vert \phi_{s_k}\rangle$. Correspondingly, each triplet of arguments summed in (\ref{eq:phi2'}) is equal to the argument of the weak value $\langle \phi_{s_k}\vert \hat{\Pi}_{r}\vert\phi_{i_k}\rangle\langle \phi_{s_k}\vert \phi_{i_k}\rangle^{-1}$ of a qubit projector on the state $\vert \phi_{r}\rangle$. Using our results (\ref{eq:argprojwvalqubit}) on qubits, we find thus that
\begin{equation}\label{eq:phasemodvalNlevelfinal}
\arg \Lambda_{r,m}^{\alpha,\beta}=\beta-\alpha\frac{N-1}{2}\Lambda_r-\frac{1}{2}\sum_{k=1}^{N-1}(\Omega_{i_k s_k f}+\Omega_{ i_k r s_k})\,,
\end{equation}
where the geometrical component can be recast as $\Omega_{i_k r s_k f}=\Omega_{i_k s_k f}+\Omega_{ i_k r s_k}$ according to (\ref{eq:quadto2trig}). When the operator $\Lambda_{r,m}^{\alpha,\beta}$ is a spin operator, the associated unitary operator $ \hat{U}_{\Lambda_{r}}^{\alpha,\beta}$ is rotation operator that corresponds to a rotation of an angle $\alpha$ in physical space and to a rotation of an angle $\frac{N-1}{2}\alpha$ in the Hilbert space of the $N$-level system. Thus, it rotates the initial vectors $\vec{i}_k$ around the axis $\vec{r}$ by an angle $\alpha$ until they reach the vectors $\vec{s}_k$. In that case, the solid angle $\Omega_{i_k r s_k f}$ is given in closed form by (\ref{eq:gmphasemodularirsf}) so that it is not necessary to know the individual vectors $\vec{s}_k$ to determine the geometrical phase. When $ \hat{U}_{\Lambda_{r}}^{\alpha,\beta}$ is not a spatial rotation operator, the expression of $\vec{s}_k$ as a function of $\vec{i}_k$, $\vec{r}$ and $\alpha$ is a priori not kown, so that the phase of the modular value should be evaluated through (\ref{eq:phasemodvalNlevelfinal}), using the general formula (\ref{eq:argprojwvalqubit}) to calculate each solid angle.

\section{Majorana representation for an arbitrary state}
\label{appendix:singularities}

After application of the unitary operators, an arbitrary initial state can be written as $|\psi_{i}''\rangle=(e^{j\chi_{1}}\cos\epsilon\sin\theta,e^{j\chi_{2}}\sin\epsilon\sin\theta,\cos\theta)^{T}$. Its Majorana polynomial is given by  
\begin{equation}
z^{2}-\sqrt{2}\sin\epsilon\tan\theta\,e^{j\chi_{2}}\,z + \cos\epsilon \tan\theta\, e^{j\chi_{1}}=0\,, \label{eq:MajoranaPolynimial}
\end{equation}
where the roots $z_{1,2}=\tan(\frac{\beta_{1,2}}{2})\,e^{j\alpha_{1,2}}$ provide the coefficients of the qubits states $ \cos \frac{\alpha_{1,2}}{2} |0\rangle\ + e^{j \beta_{1,2}} \sin \frac{\alpha_{1,2}}{2} |1\rangle$. The solutions are given by
\begin{eqnarray}
\alpha_{1,2}=\frac{\chi_{1}}{2} \pm(-1)^{k}\,\arccos\left(\frac{\sqrt{2}\sin\epsilon \tan\theta}{S}\cos\tilde{\chi}\right)\,,\nonumber \\ 
\beta_{1,2}=2\arctan\left(\frac{S\pm\sqrt{S^2-4\cos\epsilon \tan\theta}}{2}\right)\:,\label{eq:alphabeta}
\end{eqnarray}  
where we defined $\tilde{\chi}=\frac{2\chi_{2}-\chi_{1}}{2}$ and $S=\sqrt{2\cos\epsilon \tan\theta + \sin^{2}\epsilon \tan^{2}\theta + \sqrt{\rho}\:\:}$ with $\rho=4\cos^{2}\epsilon \tan^{2}\theta + \sin^{4}\epsilon \tan^{4}\theta-4\cos\epsilon \sin^{2}\epsilon \tan^{3}\theta \cos(2\tilde{\chi})$. The parameter $k$ in the relation of $\alpha_{1,2}$ is zero if the condition $0\leq\tilde{\chi}<\pi$ is satisfied, and equals one if $\pi\leq\tilde{\chi}< 2\pi$.

\section{Orthogonal-state decomposition}
\label{Orthogonal-states decomposition}
The two vectors $|\Psi_{1}''\rangle$ and $|\Psi_{3}''\rangle$ are in the subspace spanned by $|\phi_{r}\rangle|\phi_{r}\rangle$ and $|\phi_{-r}\rangle|\phi_{-r}\rangle$.
The qubit states are
\begin{eqnarray}
|\phi^{(1,2)}_{m}\rangle=\left(\frac{1 \mp \sqrt{2\sqrt{3}-3}}{\sqrt{3}}\right)^{\frac{1}{2}}|0\rangle \pm \left(1-\frac{1 \pm\sqrt{2\sqrt{3}-3}}{\sqrt{3}}\right)^{\frac{1}{2}}|1\rangle\:, \\
|\phi^{(1,2)}_{n}\rangle=\sqrt{\frac{3-\sqrt{3}}{3}}\,|0\rangle + 3^{-\frac{1}{4}} \,e^{\mp i\,\phi_{n}} |1\rangle\:, \label{eq:TBP StateApproach}
\end{eqnarray}   
with $\phi_n=\arctan \sqrt{9+6 \sqrt{3}}$. To find the expressions (\ref{eq:3box-box1}) and (\ref{eq:3box-box3}) associated with boxes one and three, the procedure is to construct the symmetrized states according to the first equation of formula (\ref{eq:AlgStr}), but after making a basis change from $\{\vert 0 \rangle, \vert 1 \rangle\}$ to $\{\vert \phi_r \rangle, \vert \phi_{-r} \rangle\}$. This gives the qutrit states $(\frac{\sqrt{3}}{2},0,\frac{1}{2})^T$ and $(-\frac{1}{2},0,\frac{\sqrt{3}}{2})^T$, which as expected have a nul projection on the state $\frac{1}{\sqrt{2}}(|\phi_{r}\rangle|\phi_{-r}\rangle+|\phi_{-r}\rangle|\phi_{r}\rangle)$.

\end{document}